\documentclass[preprint,aps,amsmath]{revtex4}
\usepackage{graphicx}
\usepackage{subfig}
\usepackage{color}

\begin{document}
\preprint{}

\title{Correlation length scalings in fusion\\ edge plasma turbulence computations}

\author{S. Konzett$^1$, D. Reiser$^2$, A. Kendl$^1$}
\affiliation{1) Institut f\"ur Ionenphysik und Angewandte Physik, Association
  Euratom-\"OAW, Universit\"at Innsbruck, Austria\\
2) Institut f\"ur Energie- und Klimaforschung - Plasmaphysik, Forschungszentrum
  J\"ulich, Association Euratom-FZJ, Germany \vspace{1cm}
}

\begin{abstract}
The effect of changes in plasma parameters, that are characteristic near or at
an L-H transition in fusion edge plasmas, on fluctuation correlation lengths
are analysed by means of drift-Alfv\'en turbulence computations.
Scalings by density gradient length, collisionality, plasma
beta, and by an imposed shear flow are considered.
It is found that strongly sheared flows lead to the appearence of
long-range correlations in electrostatic potential fluctuations parallel and
perpendicular to the magnetic field.

\vspace{9cm}

{\sl This is the preprint version of a manuscript submitted to Plasma Physics
  and Controlled Fusion.}
\end{abstract}

\maketitle

\section{Introduction}

The interplay between long-range correlations of turbulent fluctuations,
radial electric fields and edge bifurcations in fusion plasmas has received
recent interest in the form of various experimental studies
\cite{Moyer01,Pedrosa08,Xu09,Manz10,Wilcox11,Xu11,Silva11,Stroth11}.
This interest is motivated by a missing mechanism behind the formation
of edge transport barriers at the transition from L- to H-mode plasma states.
The central link between appearence of radial electric field $E_r$ and
associated sheared $E_r$$\times$$B$ flows to suppression of small-scale
turbulence, the reduction in turbulent transport, and a steepening of the
pedestal profile, is generally accepted~\cite{Stroth11}. 
However, the causal chain of mechanisms behind this transport barrier
formation is as yet unclear. 

It has been speculated that turbulence generated zonal
flows could be able to trigger the mean shear flow bifurcation.
Long-range correlations in turbulent fluctuations have been associated with
enhanced zonal flow activity. 
In L-mode experiments with imposed shear flow an increase in correlation
length of the fluctuating electrostatic potential~\cite{Pedrosa08} and 
density~\cite{Manz10} has been found along and across magnetic field lines. 

The influence of single possible players behind the formation of long-range
correlations can not always easily be determined by experiments, but may be
straightforwardly studied with numerical simulation.
In this work, local drift-Alfv\`en flux-tube turbulence computations are
applied to analyse correlation statistics for various L-mode parameters in
scalings that are characteristic for the approach to the H-mode. 
In particular, scaling effects by the background density gradient length, the
collisionality, the plasma beta, an imposed $E$$\times$$B$ shear flow and
zonal flows on correlation statistics are studied.
It is found that only strong imposed shear flows are able to generate
significant long-range correlations in these simulations.

The work is organized as follows:
In Sec.~II the numerical model and reference parameters are described. In
Sec.~III the evaluation of correlation functions from fluctuating simulated
quantities is reviewed. In Sec.~IV the individual scaling relations
are analyzed, followed by the conclusions in Sec.~V.

\section{Numerical model: drift-Alfv\`en turbulence}

The four-field drift-Alfv\'en fluid model \cite{Scott97}
for electromagnetic fusion edge plasma turbulence is solved numerically using the local
flux-tube code ATTEMPT \cite{Reiser09}. 
The model describes the evolution of fluctuations of the electrostatic
potential $\phi$, particle density $n$, vector potential $A_{\|}$ and 
parallel ion velocity $u$:

\begin{eqnarray}
{dn \over dt} &=& {1 \over e} \nabla_{\|}\,J - n \nabla_{\|} u - n {\cal K}(\phi)
 + {T_{e} \over e}\,{\cal K}(n) \\ 
 {nm_{i} \over B_{0}^{2}}\,{dw \over dt} &=& \nabla_{\|} J + T_{e}\,{\cal K}(n) \\
 {\partial A \over \partial t} + {m_{e} \over e^{2} n}\,{d J \over d t} &=&
 {T_{e} \over e n}\,\nabla_{\|} (n-\phi) - \eta_{\|}\,J \\     
 n\,m_{i}\,{d u \over dt} &=& -T_{e}\,\nabla_{\|} n  
\end{eqnarray}

This set of equations is coupled to the solution of Poisson's and Ampere's 
equations for the vorticity $w$ and the vector potential $A_{||}$:

\begin{equation}
 w = \nabla_{\perp}^{2} \phi \qquad \mbox{and} \qquad \mu_{0}\,J = -
 \nabla_{\perp}^{2} A_{\|}.
\end{equation}

Operator abbreviations have been introduced as follows:

\begin{eqnarray}
&&{d \over dt} = {\partial \over \partial t} + \mathbf{v}_{E} \cdot \nabla 
\qquad \mbox{with} \qquad
  \mathbf{v}_{E} \cdot \nabla  = {\mathbf{B} \over B^{2}}\cdot(\nabla \phi \times \nabla ) \\
&&  {\cal K} = \nabla \cdot \left( {\mathbf{B}} \times \nabla  \over B^{2} \right) \\
&& \nabla_{\|}  = {\mathbf{B}\over B} \cdot \nabla - {\mathbf{B} \over B^{2}}
 \cdot (\nabla A_{\|} \times \nabla ) \\ 
&&   \nabla_{\perp}^{2}  = \nabla^{2}  - \nabla \cdot {\mathbf{B}\over B}{\mathbf{B}\over B} \cdot \nabla 
\end{eqnarray}

A static toroidal equilibrium background magnetic field $\mathbf{B}$ is assumed.
The model describes nonlinear electromagnetic $E$$\times$$B$ drift motions of electrons of
mass $m_{e}$ and ions of mass $m_{i}$ with charges $q=\pm e$. 
The ion and electron particle densities are equal, $n_{e}=n_{i}=n$, obeying
quasi-neutrality.   
Ions are cold and electrons have the constant temperature $T_e$, and the 
electron-ion collision frequency is $\nu_{e}$.

A local approximation is applied, where the density gradient is linear and
constant in time with $L_n^{-1}= |\nabla \ln n_0|$ with axisymmetric
background density $n_0$ and the density $n = n_0 + \tilde n$ splitted into a
static and a fluctuating part. In the following the tilde on the fluctuating
density will be avoided for better readability. 
A partially field-aligned flux-tube coordinate system $(x, y, z)$ is introduced
and the standard drift normalisation is applied, which are described in detail
in the Appendix of Ref.~\cite{Reiser09}, where the coordinates 
$(x, y, z)$ are denoted by $(\chi,\eta,\sigma)$. 
Parallel derivatives (in $z$ direction) are normalized with respect to the
parallel connection length $L_{\|} = 2 \pi q R_{0}$, perpendicular derivatives
with $L_{\perp}$, and time scales with $(c_s / L_{\perp})$.
The density gradient length then enters via $\lambda_n = {L_{\perp} / L_n} =
|\partial_x \ln n_0|$. 
The normalized set of equations is: 

\begin{eqnarray} 
 {\partial n \over \partial t} &=& - \{ \phi,n \} - \lambda_n\,\partial_{y}
 \phi - {\cal K}(\phi-n) + \nabla_{\|} (J-u) \\
 {\partial w \over \partial t} &=& - \{ \phi,w \} + \nabla_{\|} J + {\cal K}(n)  \\
\hat{\beta}\,{\partial A_{\|} \over \partial t} + \hat{\mu}\,{\partial J \over
  \partial t} &=& -\{ \phi,J \} + \nabla_{\|} (n-\phi) - \hat{C}\,J \\
\hat{\epsilon}\,{\partial u \over \partial t} &=& -\{ \phi,u \} - \nabla_{\|} n 
\end{eqnarray}

In a large aspect ratio circular flux-tube geometry, the Poisson bracket is
$ \{ f,g \} = \partial_x f \; \partial_y g  - \partial_y f \; \partial_x g$, 
the curvature operator is ${\cal K}(f) = -\omega_{B}\;[ \cos(s)\; \partial_x f
  + \sin(s)\;\partial_y f ]$, 
the parallel derivative $\nabla_{\|} f = \partial_z f  - \hat \beta \; \{ A,f \}$,
and the Laplacian becomes 
$ \nabla_{\perp}^{2}  = \partial^{2}_x   + \partial^{2}_y $.
The numerical methods using a higher-order Adams-Bashforth / Arakawa scheme
are detailed in Ref.~\cite{Reiser09}.

Simulations have been performed using reference edge parameters typical of the
TEXTOR experiment, with major radius $R_{0}=1.74$~m, minor radius $a=0.5$~m,
electron temperature $T_{e}=51.8$~eV, magnetic field strength $B_{0}=1.0$~T, 
plasma density $n_{0}=5.6 \cdot 10^{18}$~m$^{-3}$, a background density
gradient reference scale $L_{\perp} = 3.54$~cm, and a parallel connection scale
 $L_{||} = q R_0 = 465$~cm with $q=2.66$ and $\hat{s} =
  (a/q)(\partial q / \partial r) = 1$.
Conversion to dimensionless model parameters \cite{Reiser09} gives 
a parallel to  perpendicular scale ratio  $\hat{\varepsilon}=(L_{\|}/L_{\perp})^{2}=17226$, 
collisionality $\hat{C}= \hat{\mu} L_{\perp}/c_{s} \nu_{e}/1.96=1.0$, 
beta $\hat{\beta} = \mu_{0} n_{0} T_{e} / B_{0}^{2}=1.0$,
mass ratio $\hat{\mu}=\hat{\varepsilon} m_{e}/m_{i}=4.69$, 
and curvature scale $\omega_{B} = 2 L_{\perp}/R_{0}=0.046$.
The numerical grid resolution is set to  $L_x \times L_y \times
  L_z = 64 \rho_s \times 256 \rho_s \times 16 L_{||}$.

The magnitude of the dimensionless parameters $\hat{C}$, $\hat{\beta}$ and
$\hat{\mu}$ in the order of unity is typical for many fusion edge plasmas,
including larger tokamaks and some stellarator experiments. 
The simulations and results are therefore rather generic and not restricted in
their applicability on a specific tokamak configuration like TEXTOR.

These nominal values are varied in the simulations to account for changes
in pedestal parameters related to the approach towards H-mode conditions. 
The simulations are run into a fully developed saturated turbulent state,
where time series of density and potential fluctuations are recorded at
several ``probe'' position, and are subjected to a correlation length analysis. 

\section{Evaluation of correlation functions}

In this section correlation functions used in the following analysis are
reviewed.
The auto-correlation (AC) function $\gamma_{auto}$ of a fluctuation signal
$f(t)$ is defined as \cite{Dunn05}:
\begin{eqnarray} 
 \gamma_{auto}(\tau) = {1 \over T} {\displaystyle \sum\limits_{t=0}^{T}
   f(t+\tau)\;f(t) \over f(t)^{2}}.
\label{eq:autocorr}
\end{eqnarray}
A windowed AC analysis on the computed time series is performed by shifting 
a slice of the data $f$ of size $\Delta T$ by $\delta_{T}$ for every step.
Here we use $\delta_{T} = 0.3~L_{\perp}/c_{s}$ and $\Delta T = 60~L_{\perp} / c_{s}$.
The AC function is evaluated in the interval
 $[t_{i},t_{i}+\Delta T]$ with $t_{0}=0$, $t_{1}=\delta_{T}$, $t_{i}= i\;\delta_{T}$:
\begin{equation}
 \gamma_{auto}(\tau,t_{i}) = {1 \over \Delta T}{ \displaystyle \sum\limits_{t=i\;\delta_{T}}^{\Delta
     T+i\;\delta_{T}} f(t+\tau)\;f(t) \over f(t)^{2}}  
 \label{eq:autocorrtime2}
\end{equation}
By evaluation at every time step a time series of the self correlation time
$\tau_{AC}(t_{i})$ defined by $\gamma_{auto}(\tau_{AC}(t_{i}),t_{i}) \equiv 0.5$ is obtained.
The statistical properties of $\tau_{AC}(t_{i})$ are
displayed using probability density functions (PDF):

\begin{equation}
 P(\tau_{AC})= P( t_{b-1} < \tau_{AC} <  t_{b}=t_{b-1}+dt_{b} ) = { 1 \over N}
 \sum_{t_{b-1} < \tau_{AC} <  t_{b}} \delta(t-\tau_{AC})
\end{equation}

where $N$ is the length of the fluctuation time series $f(t)$,
$t_{b}$ is the position of a bin center, with $dt_{b}=(\max{\tau_{AC}}-\min{\tau_{AC}})/N_b$ the
width of a bin and $N_b$ the number of bins used. $P$ gives the probability of finding
the auto correlation time $\tau_{AC}$ in the fluctuation time series.

Spatial correlation lengths are analysed by means of  the cross-correlation
function of two time series $f(t)$ and $g(t)$ are fluctuation time series at two spatial positions:
\begin{equation} 
\label{eq:crosscorr}
\gamma_{gf}(\tau)  =  {\displaystyle {1 \over T} \sum\limits_{t=0}^{T} (f(t+\tau)-\bar{f})\;
   (g(t)-\bar{g}) \over \sigma(f)\;\sigma(g)} 
\end{equation}
with
\begin{equation}
\bar{f}  =  {1 \over T} \sum_{t=0}^{T} f 
\qquad \mbox{and} \qquad
\sigma(f)  =  {1 \over T} \sqrt{ \sum_{t=0}^{T} (f-\bar{f})^{2} }.
\end{equation}

To get a measure for the spatial coherence of fluctuations the cross-correlation coefficent
$CC(f,g)= \gamma_{gf}(0)$
is evaluated as the correlation function $\gamma_{gf}(\tau)$ in the limit $\tau=0$. 
The spatial correlation function $L_{CC}$ is calculated as the cross-correlation
 between a fluctuation signal $f$ taken at a probe at position $l_{0}$:$f(l_{0},t)$
 and at a spatially shifted position $l_{j}$:$f(l_{j},t)$, with
the distance between the probes $\delta l:=|l_{j}-l_{0}|$:
\begin{equation} 
\label{eq:LCC}
L_{CC} (\delta l) = CC(f(l_{0}),f(l_{0}+\delta l))
\end{equation}

A statistical description is used, where the data $f(t,l)$ is cut into pieces of length
$\Delta T$, as for the auto-correlation time above.
The correlation length (\ref{eq:LCC}) is evaluated for a time window
$[ i \delta_{T} : \Delta T+i \delta_{T} ]$. A time series
of half width times $\lambda_{l}(t_{i})$ results with 
$t_{0}=0$,$t_{1}=\delta_{T}$,$t_{i}= i \delta_{T}$:
\begin{equation} 
L_{CC} (\delta l,t_{i}) = CC(f (t_{i},l_{0}),f (t_{i},l_{0}+\delta l))
 \label{eq:autocorrtime}
\end{equation}
The correlation length $\lambda_{l}(t_{i})$ is defined
as the half width of the correlation function at $L_{CC} (\lambda_{l}(t_{i}),t_{i}) = 0.5$.
$\lambda_{l}(t_{i})$ is binned
into a histogram and normalised to one, giving a probability density function
\begin{equation} 
\label{eq:pdfcorrlength}
 P(\lambda_{l})= P( l_{b-1} < \lambda_{l} <  l_{b}=l_{b-1}+dl_{b} ) = { 1 \over N}\;
{ \displaystyle\sum\limits_{l_{b-1} < \lambda_{l} <  l_{b}} \delta(t-\tau_{AC})}
\end{equation}
where $l_{b-1},l_{b}$ are bin centers, $dl_{b}$ is the width of the bins.

\section{Scalings of correlation lengths}

When the L-H transition is approached from an L-mode state, several parameters of the
edge pedestal are changing that characteristically influence the turbulence and transport.
In the transition to the H-mode the pedestal density and temperature rise.
The collisionality $\hat{C}\sim 1/\nu_e$ is reduced as the edge temperature grows,
the plasma beta $\hat{\beta}=\hat{\epsilon} nT/\mu_{0}B_{0}^{2}$ increases 
and the density gradient length $L_{n}$ becomes smaller. 

Around the L-H transition a mean $E \times B$ flow shear layer would develop
within the edge pedestal region. As turbulence codes to date are unable to
self-consistently account for realistic H-mode shear flow development, we
model this effect by imposing a background vorticity on the turbulence.

The influence of these respective parameter scalings, which model the approach to an
H-mode state, on fluctuation correlation statistics is analysed in the following.
The reference ``probe'' position, at which the time series are recorded, is located in
the center of the computational domain, corresponding to mid-pedestal radius
($L/2$) at the torus outboard midplane position. Further analyses have been
performed for a number of radial ``probe'' position ($L/4$, $3L/8$, $5L/8$,
$3L/4$), which showed very similar results concerning scaling relations
compared to the radial reference position. Therefore only results for this
reference position are presented.

\subsection{Density gradient length scaling}

First, the steepening of the edge density gradient is modelled
by varying the density gradient length $L_{n}$ while all other parameters
remain constant at their nominal L-mode levels.
A reference simulation is performed with initial gradient length
$L_{n,0}=3.54$~cm, and four simulations with steepened gradient 
lengths $L_{\perp}/L_{n}=(1.25, 1.5, 1.75, 2.0) L_{\perp}/L_{n,0}$, corresponding to
physical gradient lengths $L_{n}=(2.99, 2.49, 2.14, 1.87)$~cm.

%%%%%%% FIGURE 1: energetics:

 \begin{figure} 
(a)~\includegraphics[width=7.0cm,height=7.0cm]{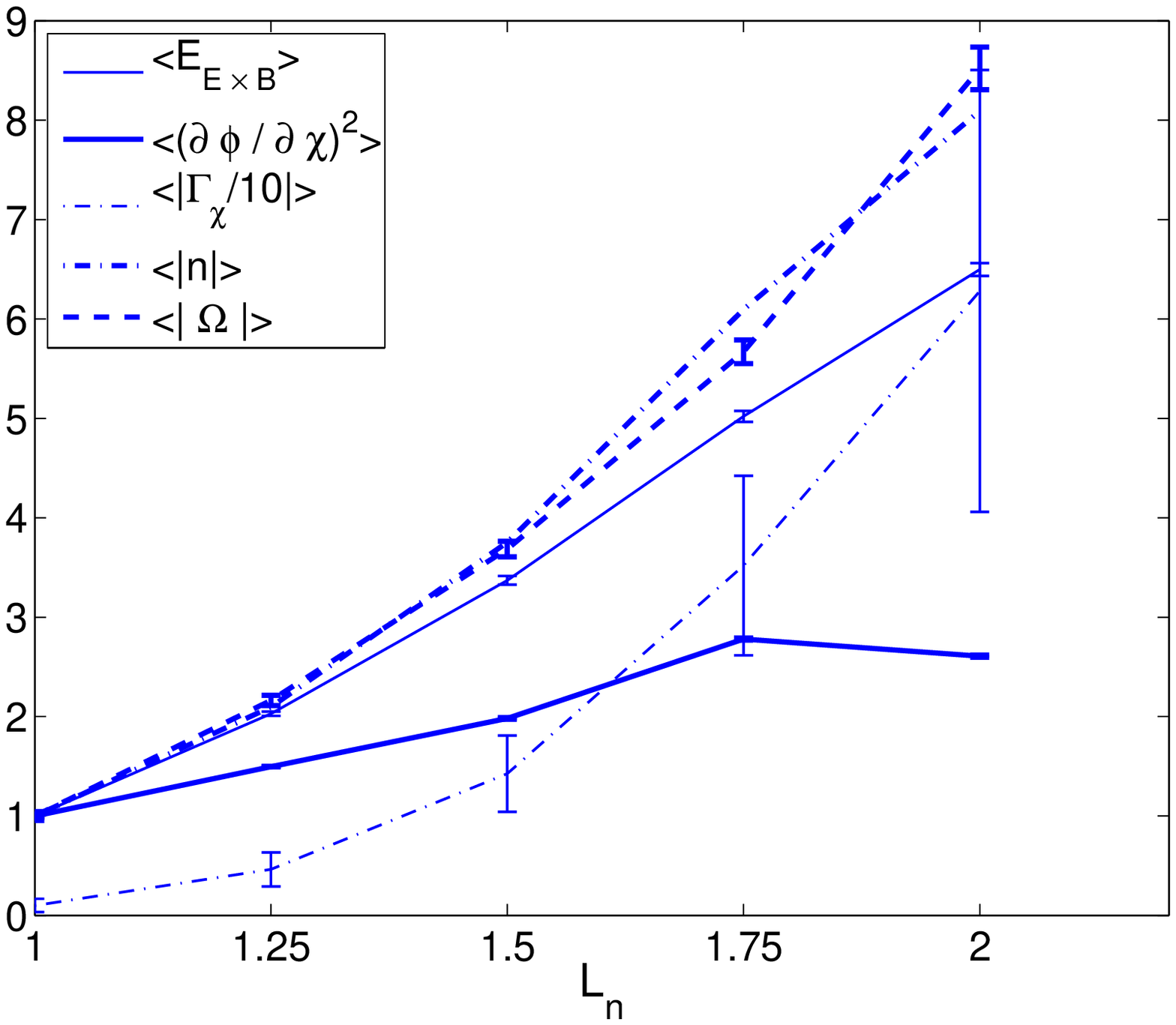}
(b)~\includegraphics[width=7.0cm,height=7.0cm]{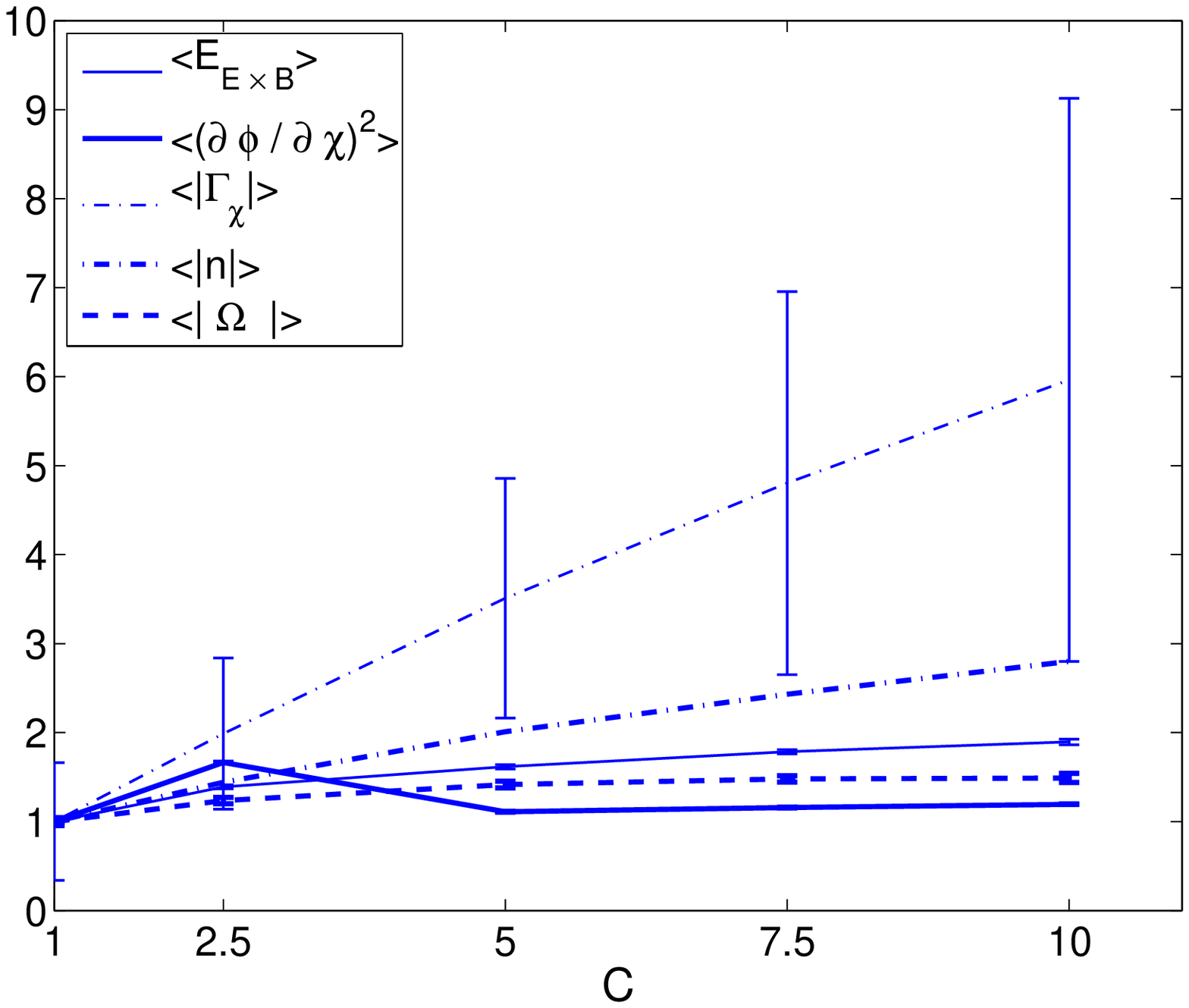}\\
(c)~\includegraphics[width=7.0cm,height=7.0cm]{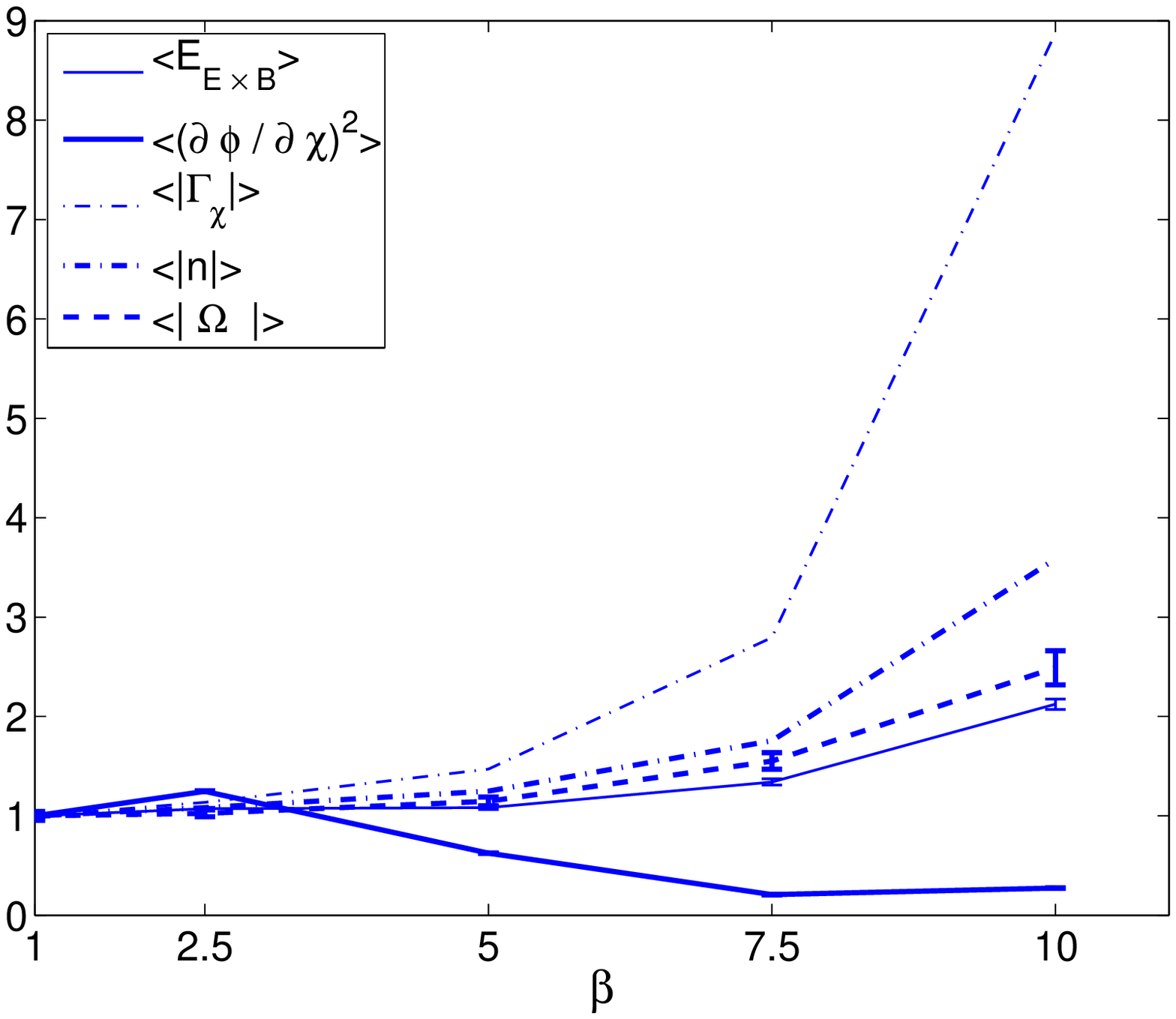}
(d)~\includegraphics[width=7.0cm,height=7.0cm]{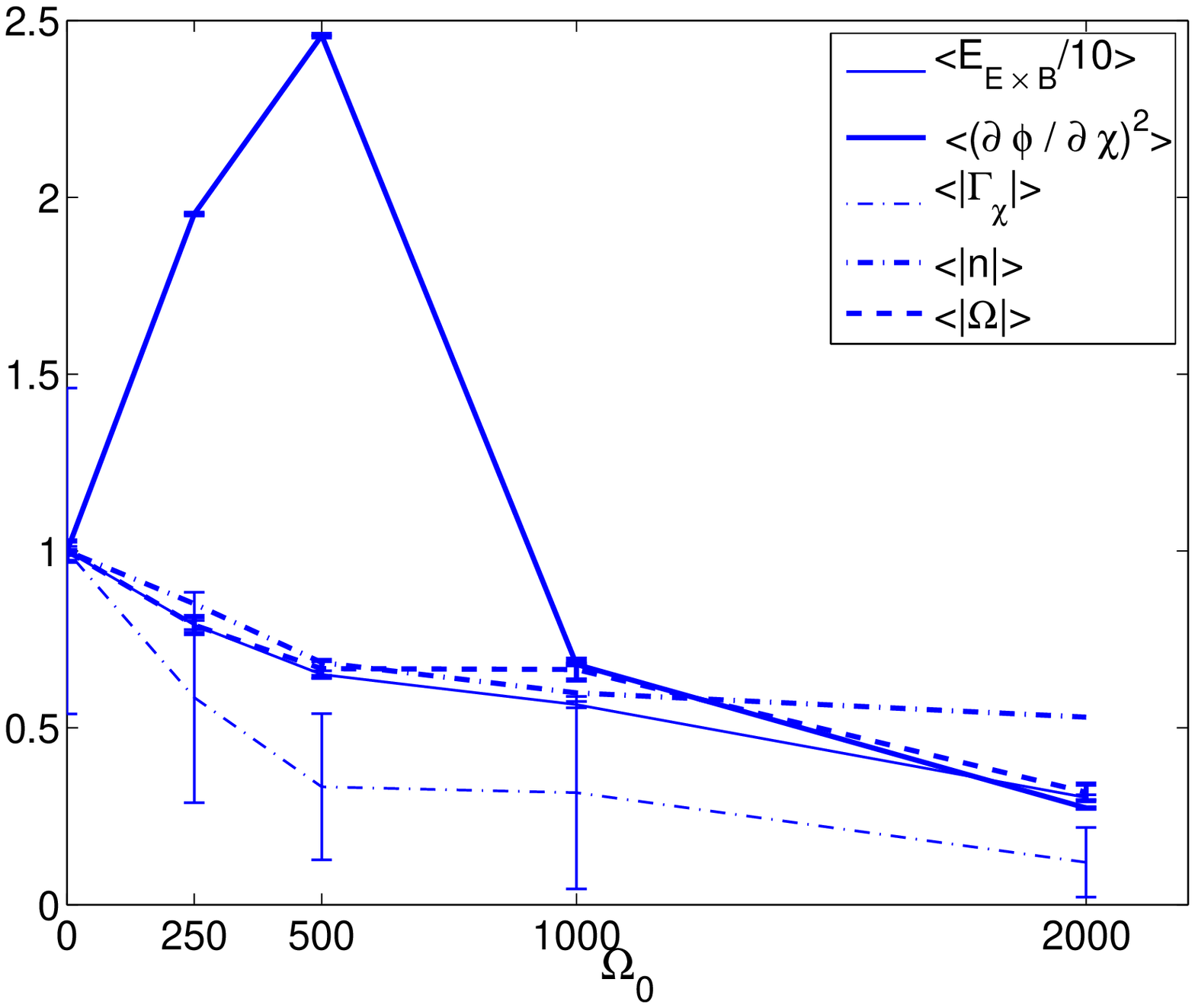}
\caption{\sl Global averages of the fluctuation kinetic energy  $\langle E_{E
\times B} \rangle= \langle \mathbf{v}_{E \times B}^{2} \rangle$ (thin solid line), 
zonal flow energy $\langle (\partial_x \phi)^{2} \rangle$ (bold solid line),
radial transport $\langle \Gamma_x \rangle= \langle v_{E \times B,x} n
\rangle$ (dash dotted line), density free energy $ \langle |n| \rangle$ 
(bold dash dotted line), and average zonal flow shear $\langle \Omega \rangle
= \langle \partial_x^2 \phi \rangle$ (bold dashed line). Scalings with
varying (a) density gradient length $L_n$, (b) collisionality $\hat C$, (c) normalised
plasma beta $\hat \beta$, (d) imposed flow shear $\Omega_0 = (\partial
v_{ZF}/\partial x)_0$. Temporal standard deviations of the fluctuating global
quantities are shown as error bars.}
\label{plot:energetics}
\end{figure}

In figure~(\ref{plot:energetics} a) global averages (over the whole
computational domain except boundary dissipation regions) 
of energetic quantities are shown. The global mean is in addition averaged
over time during the saturated turbulent phase of the simulations, and the
temporal standard deviations of the fluctuating global quantities are shown as
error bars.
The values are normalised with respect to the reference simulation $L_{n}=L_{n,0}$.

The zonal flow strength $v_{ZF}^{2}= \langle ( \partial_x \Phi_0)^2  \rangle_x$,
with $\Phi_0 (x) = \langle \phi \rangle_{y,z}$,
is doubled when the gradient is steepend corresponding to half the reference gradient length.
The zonal flow shear $\Omega_0 = \langle \partial_x^2 \Phi_0 \rangle$
increases by an order of magnitude, similar to the average free energy density
$\langle |n| \rangle$. 
The  radial density transport, defined as $\Gamma_{x}=n v_{E \times B,
x} \sim n (\partial_y \phi )$, increases by nearly two orders of magnitude.
The enhanced gradient drive is thus found to increase all turbulent activities.

\begin{figure}
% \centering
(a-1)~\includegraphics[width=4.5cm,height=4.5cm]{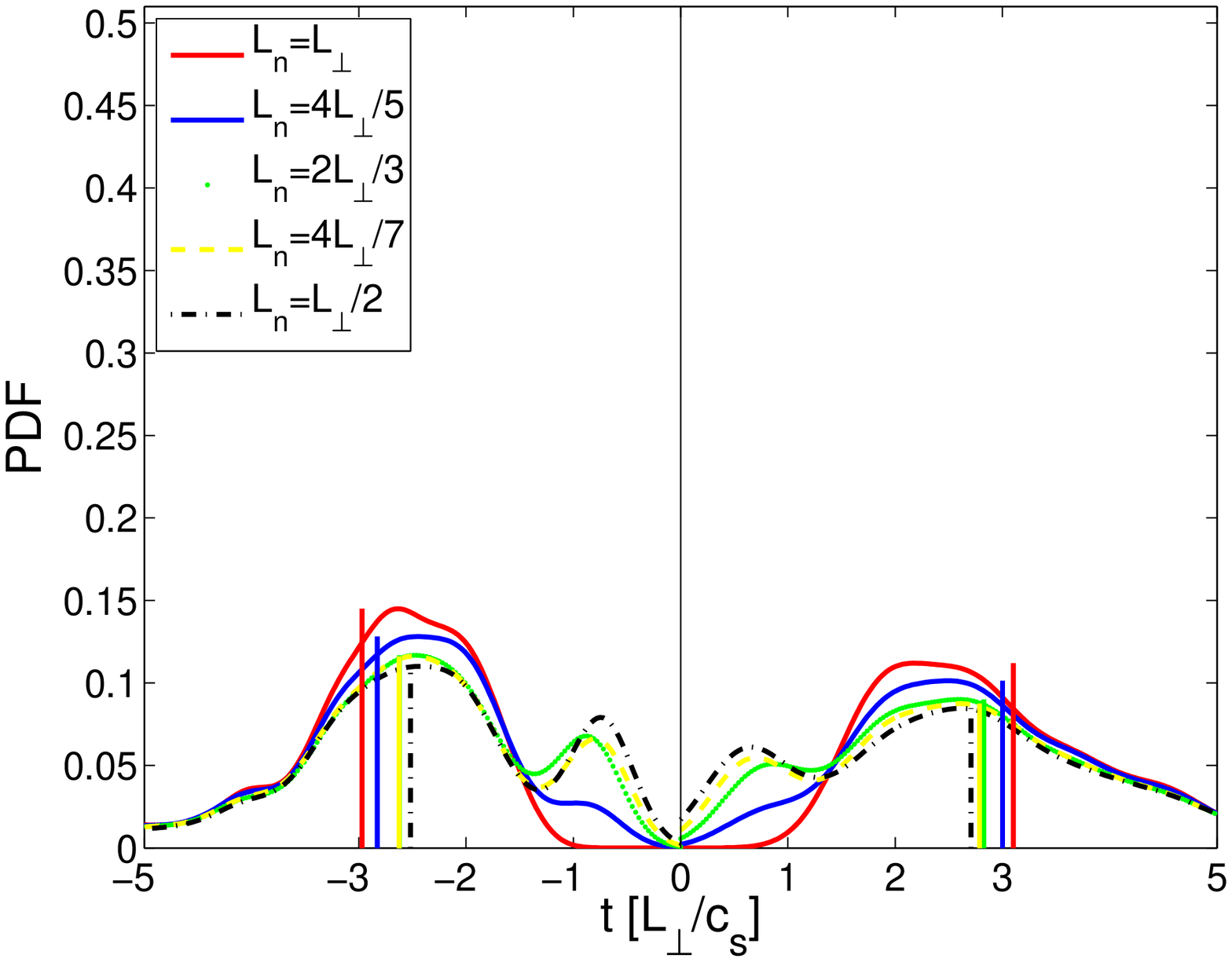}
(a-2)~\includegraphics[width=4.5cm,height=4.5cm]{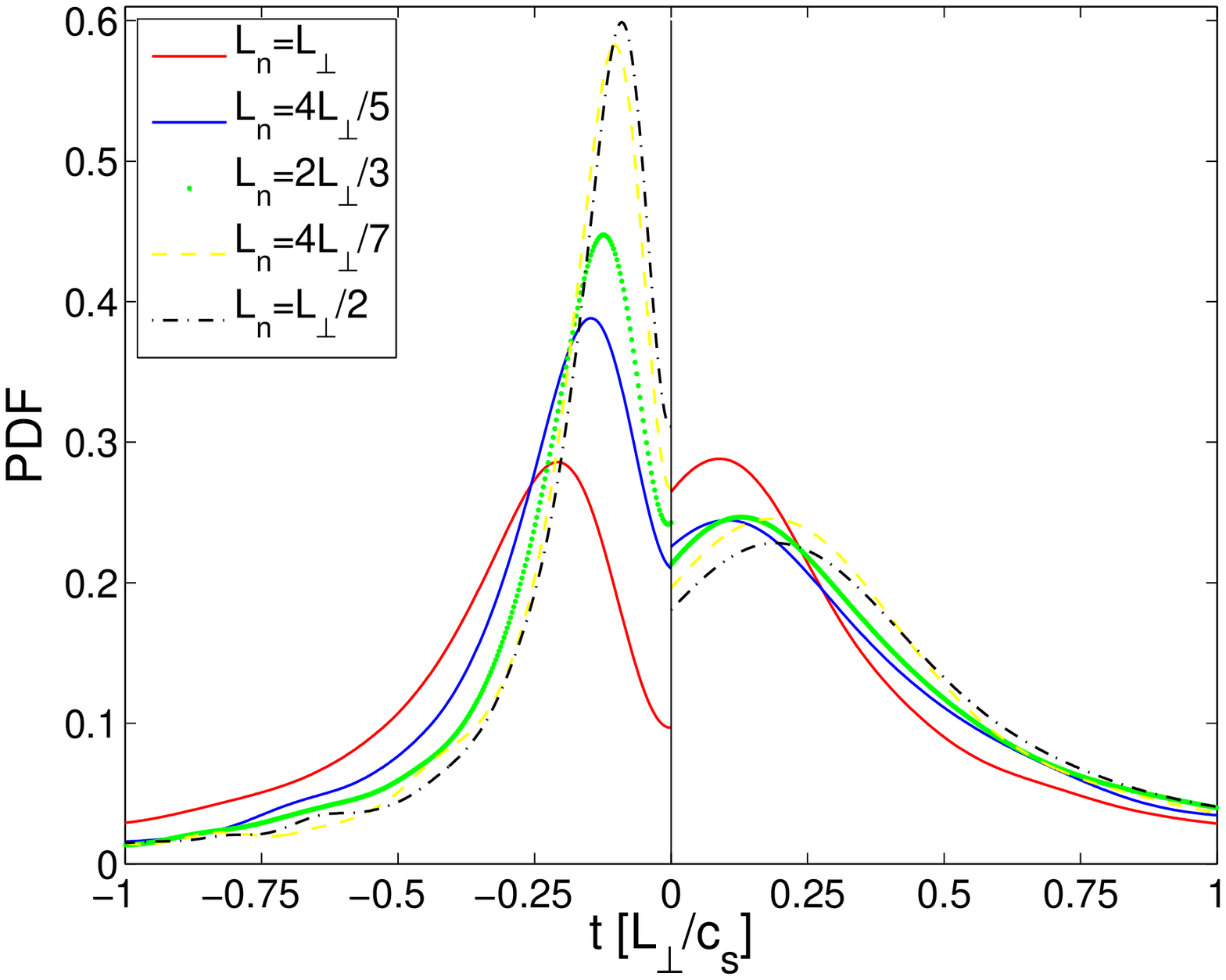}\\
(b-1)~\includegraphics[width=4.5cm,height=4.5cm]{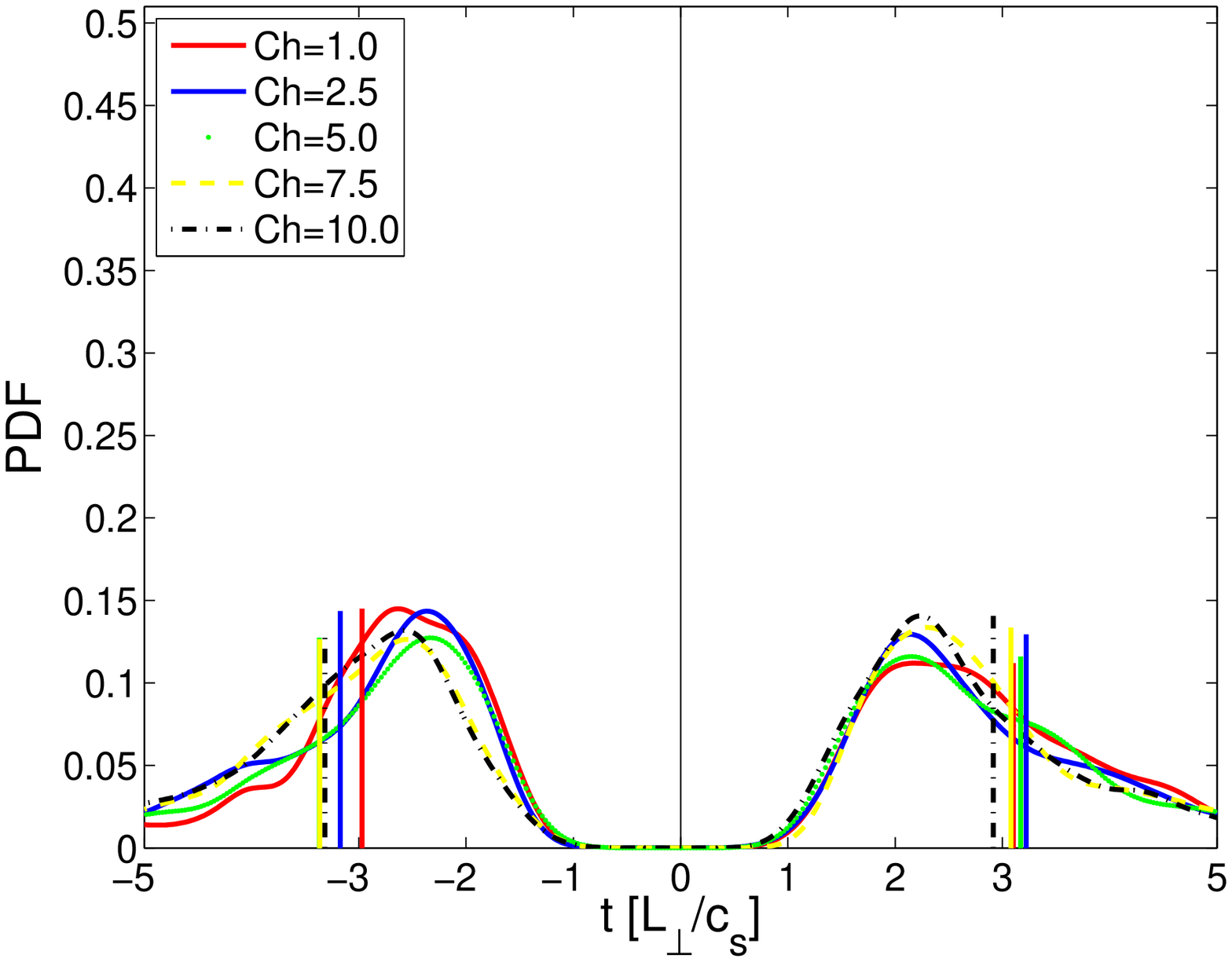}
(b-2)~\includegraphics[width=4.5cm,height=4.5cm]{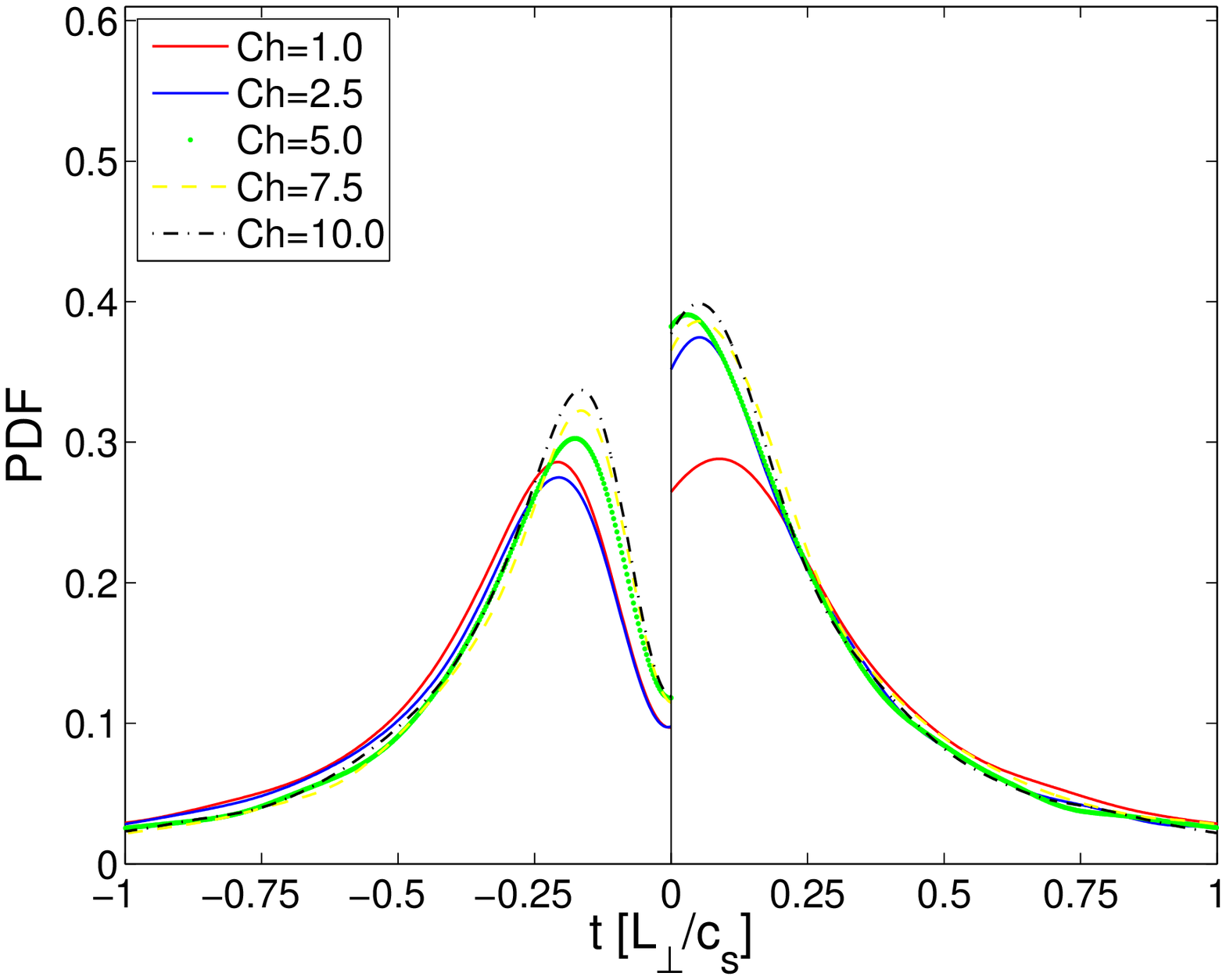}\\
(c-1)~\includegraphics[width=4.5cm,height=4.5cm]{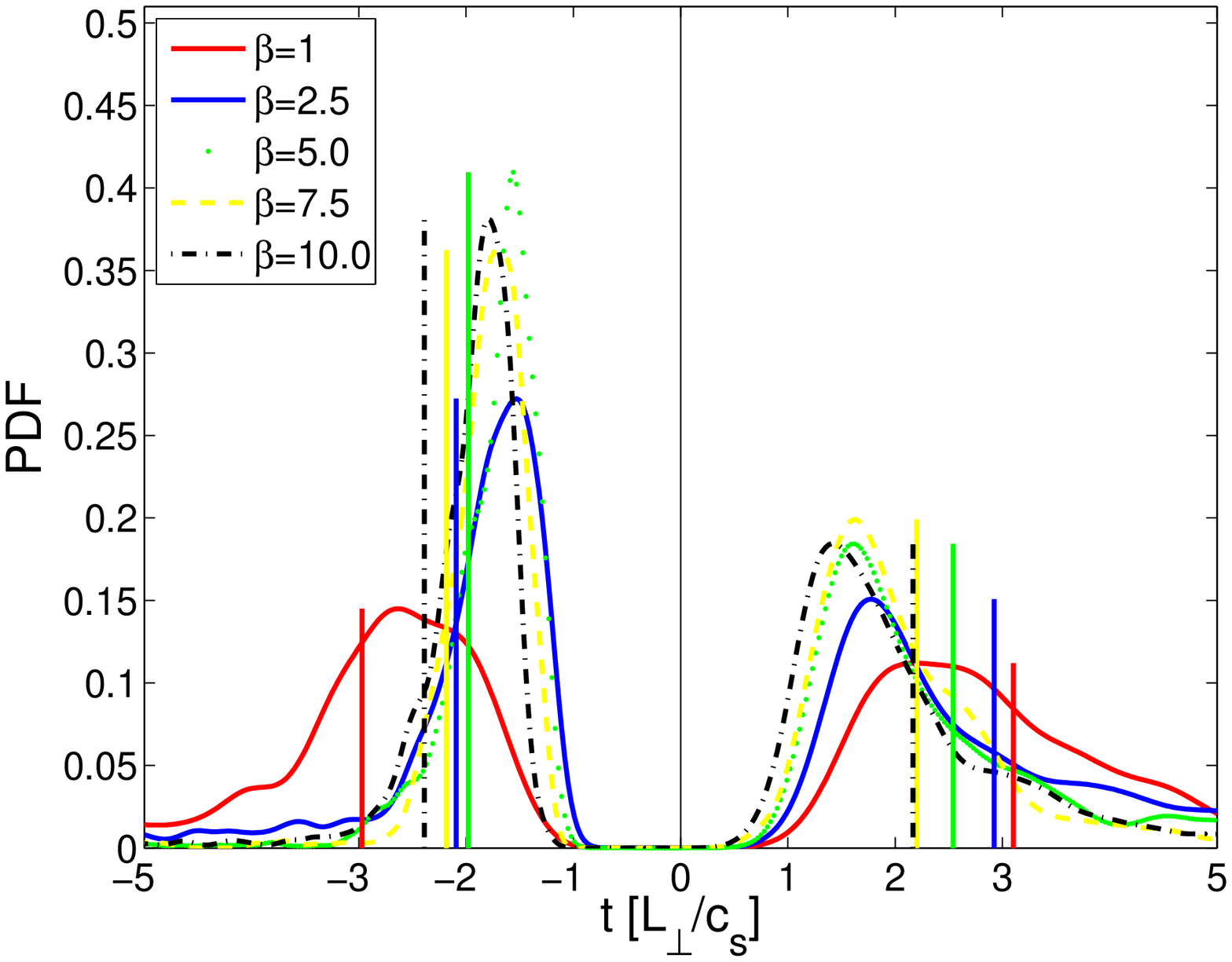}
(c-2)~\includegraphics[width=4.5cm,height=4.5cm]{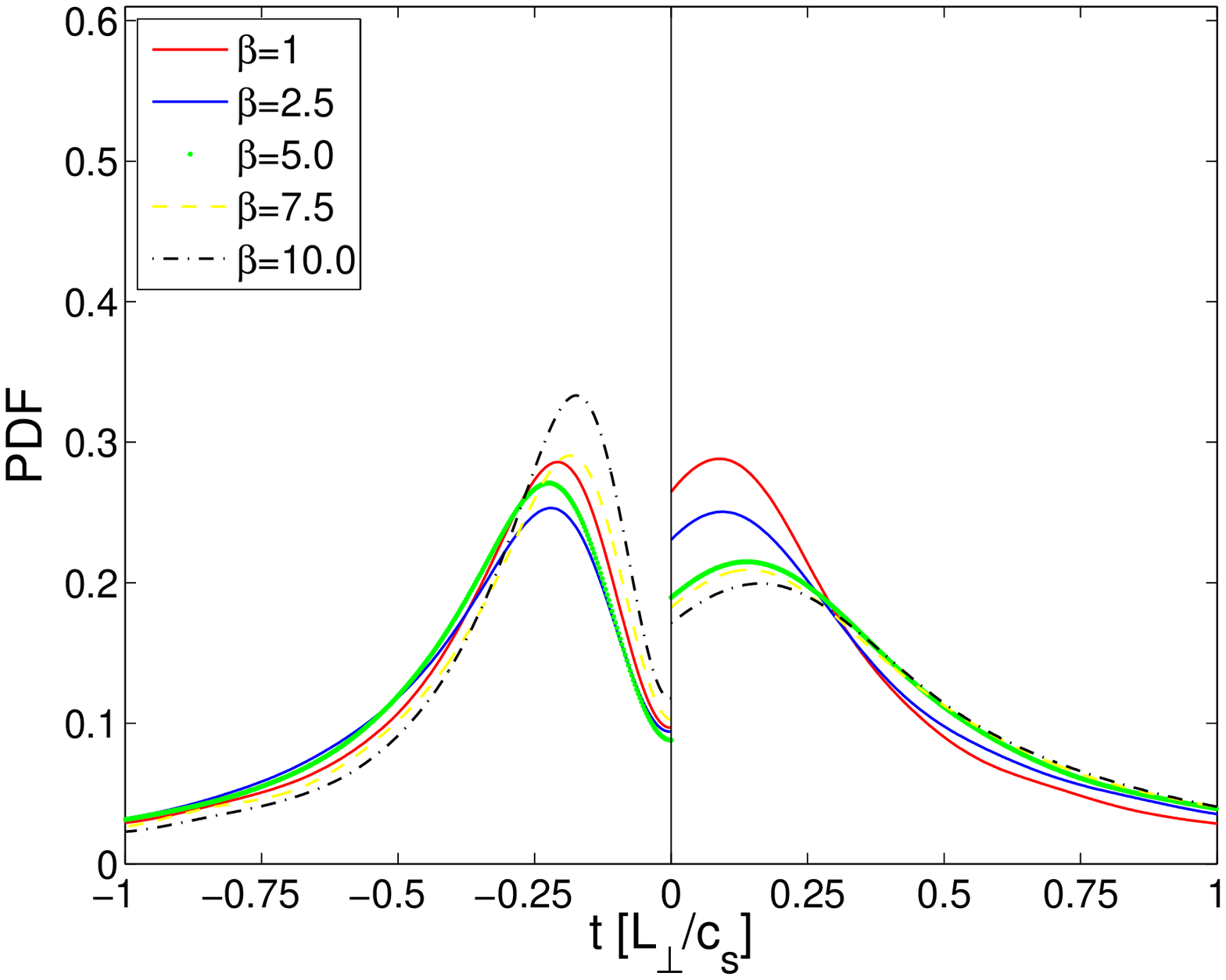}\\
(d-1)~\includegraphics[width=4.5cm,height=4.5cm]{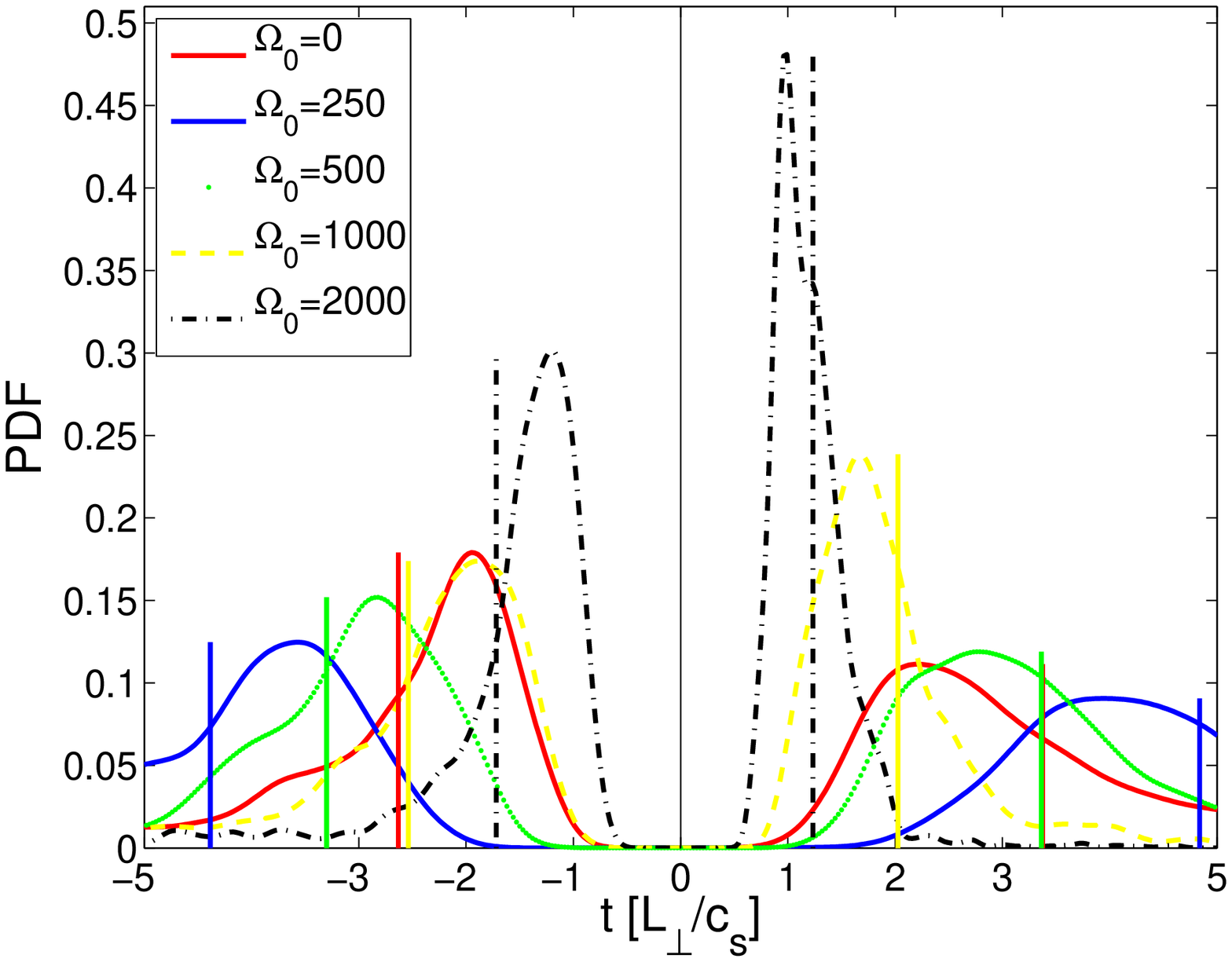}
(d-2)~\includegraphics[width=4.5cm,height=4.5cm]{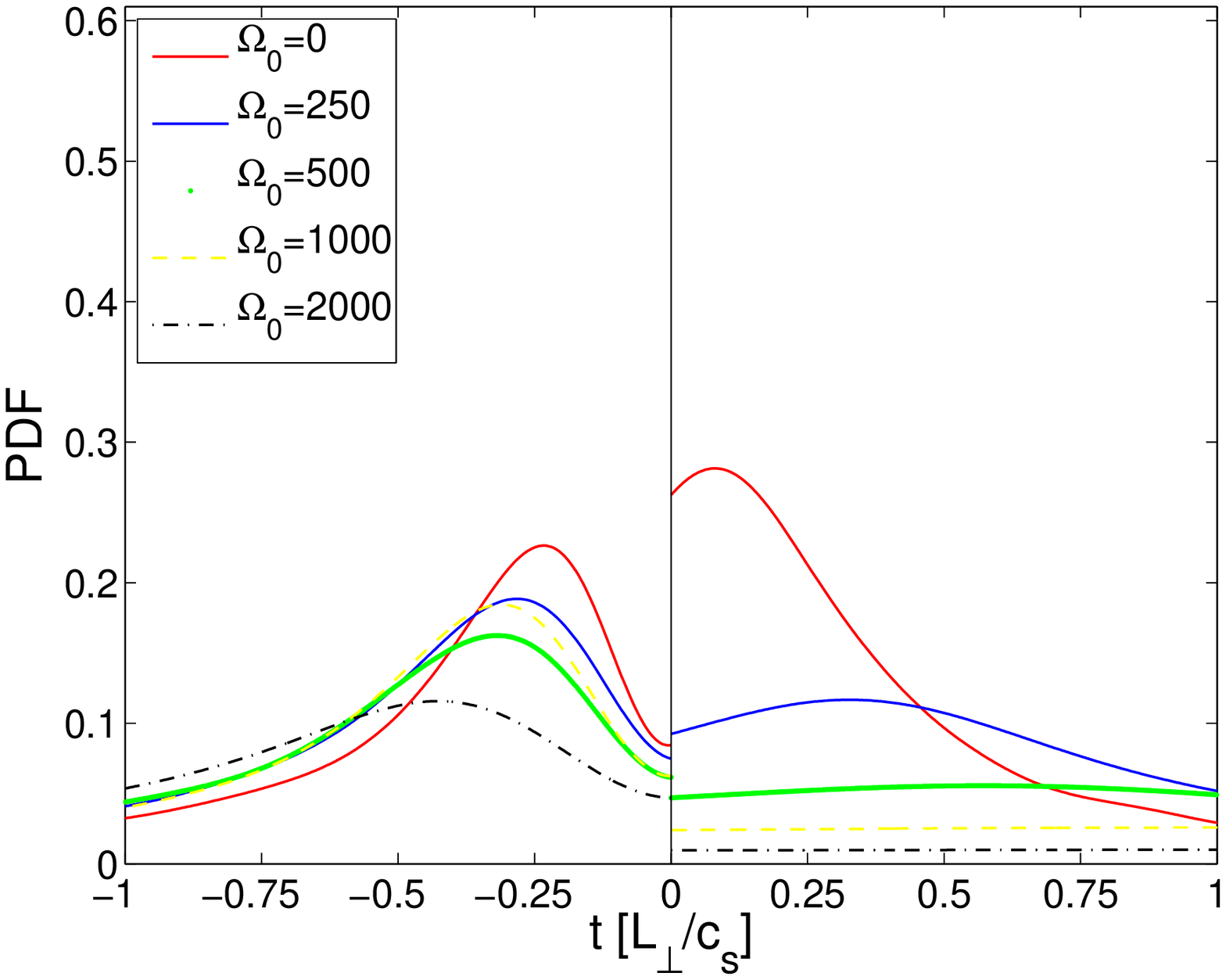}
\caption{\sl Temporal correlation scalings for (a) $L_n$, (b) $\hat C$, (c) $\hat
  \beta$, (d) $\Omega_0$: (1) auto-correlation PDFs for density (positive
  axis) and potential (negative axis);  (2) PDFs of density convection time
  scale $\tau_{conv}$ (positive axis) and $E \times B$ time scale $\tau_{E
    \times B}$ (negative axis). Mean values of AC times are drawn as vertical lines.
} 
\label{fig:tempcorr}
 \end{figure}

In figure~(\ref{fig:tempcorr} a-1) the PDFs of the fluctuation auto-correlation 
$P(\tau_{AC})$ are shown for electrostatic potential perturbations $\phi$ on
the negative axis, and density perturbations $n$ on the positive axis. 
Maxima of the auto-correlation time are found around $4~L_{\perp}/c_{s}$. 
For increasing density drive, secondary smaller peaks emerge around  $1~L_{\perp}/c_{s}$. 
The mean value of density AC is lower by around $0.5~L_{\perp}/c_{s}$ compared
to potential AC. Mean values of AC times are drawn as vertical lines.

Figure (\ref{fig:tempcorr} a-2) shows on the negative and positive axes
respectively the PDFs for two ``typical'' nonlinear time scales.

To estimate time scales for turbulent processes, we first introduce a
``convective density time scale'' $\tau_{conv}$ which should serve as a
measure for the rate of change of the density, given by the continuity equation
(which in the turbulent state is mainly determined by nonlinear convection):
$\partial_t n \sim {\bf v}_{E\times B}  \cdot \nabla n \equiv n / \tau_{conv}$ 
such that  $\tau_{conv} \equiv (n / \partial_t n) = (n / |\mathbf{v}_{E \times
  B} \cdot \nabla n| )$. 

As a second time scale, a measure of the local radial ExB velocity is
introduced through ${\bf v}_{E\times B} \sim \partial_y \phi \equiv \rho_s /
\tau_{E \times B}$, giving $\tau_{E \times B} \equiv \rho_s / \partial_y \phi$
in normalised units.

The PDF with respect to $\tau_{conv}$ is shown on the right half-space of
fig.~(\ref{fig:tempcorr} a-2): no scaling of this time scale with gradient
length is observable. 
The PDF with respect to $\tau_{E \times B}$ is shown on the left half-space of
fig.~(\ref{fig:tempcorr} a-2): increasing the gradient drive (corresponding to
smaller $L_n$) leads to a maximum of the PDF at smaller  $\tau_{E \times B}$,
which therefore can be interpreted as enhancing the radial $E \times B$ velocity.  
The correlation time of this maximum corresponds to the emerging second maximum in the auto
correlation statistics of figure (\ref{fig:tempcorr} a-1).

%%%%%%% FIGURE 2: temp corr PDFs:

\begin{figure} 
% \centering
(a-1)~\includegraphics[width=4.2cm,height=4.3cm]{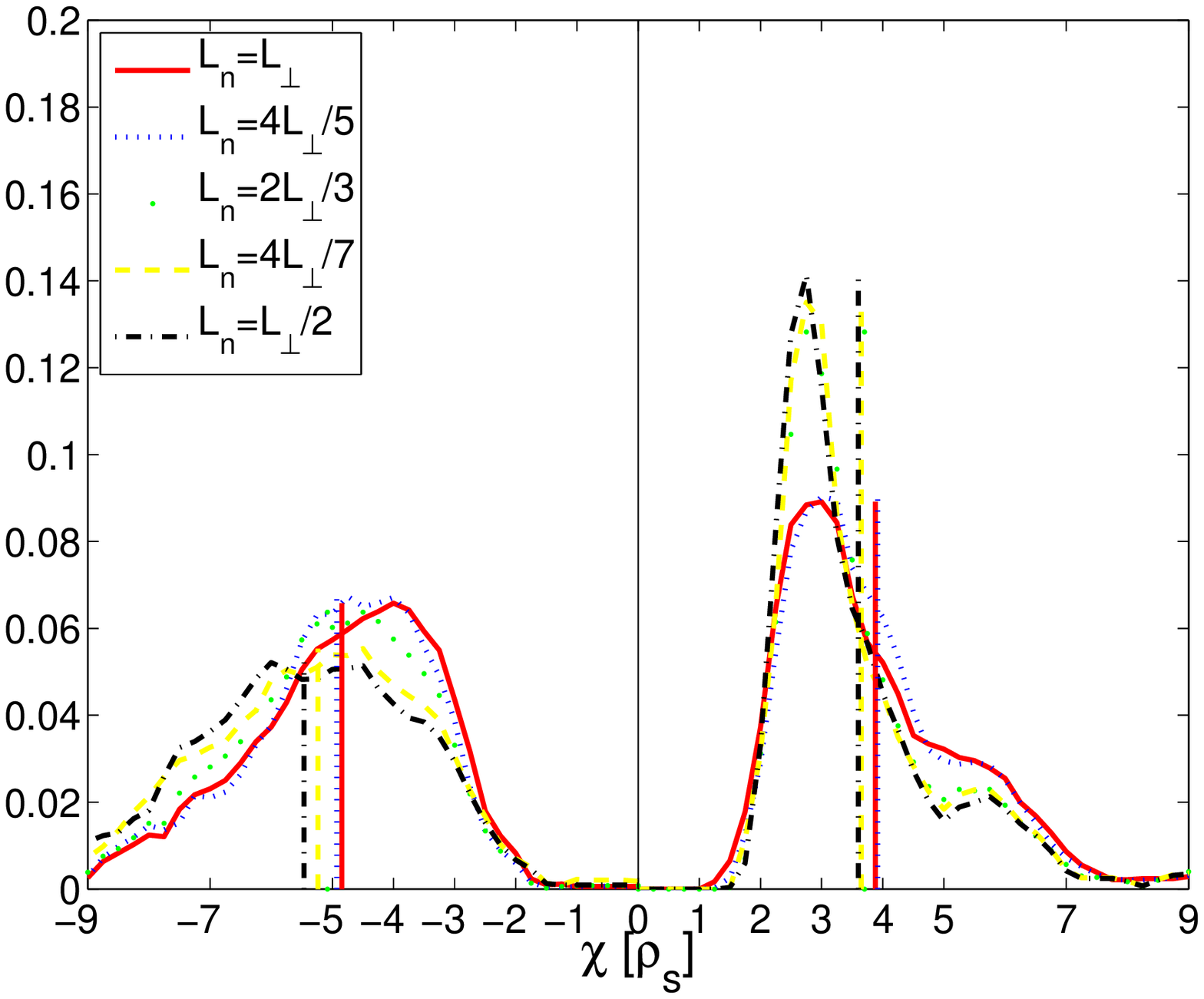}
(a-2)~\includegraphics[width=4.2cm,height=4.3cm]{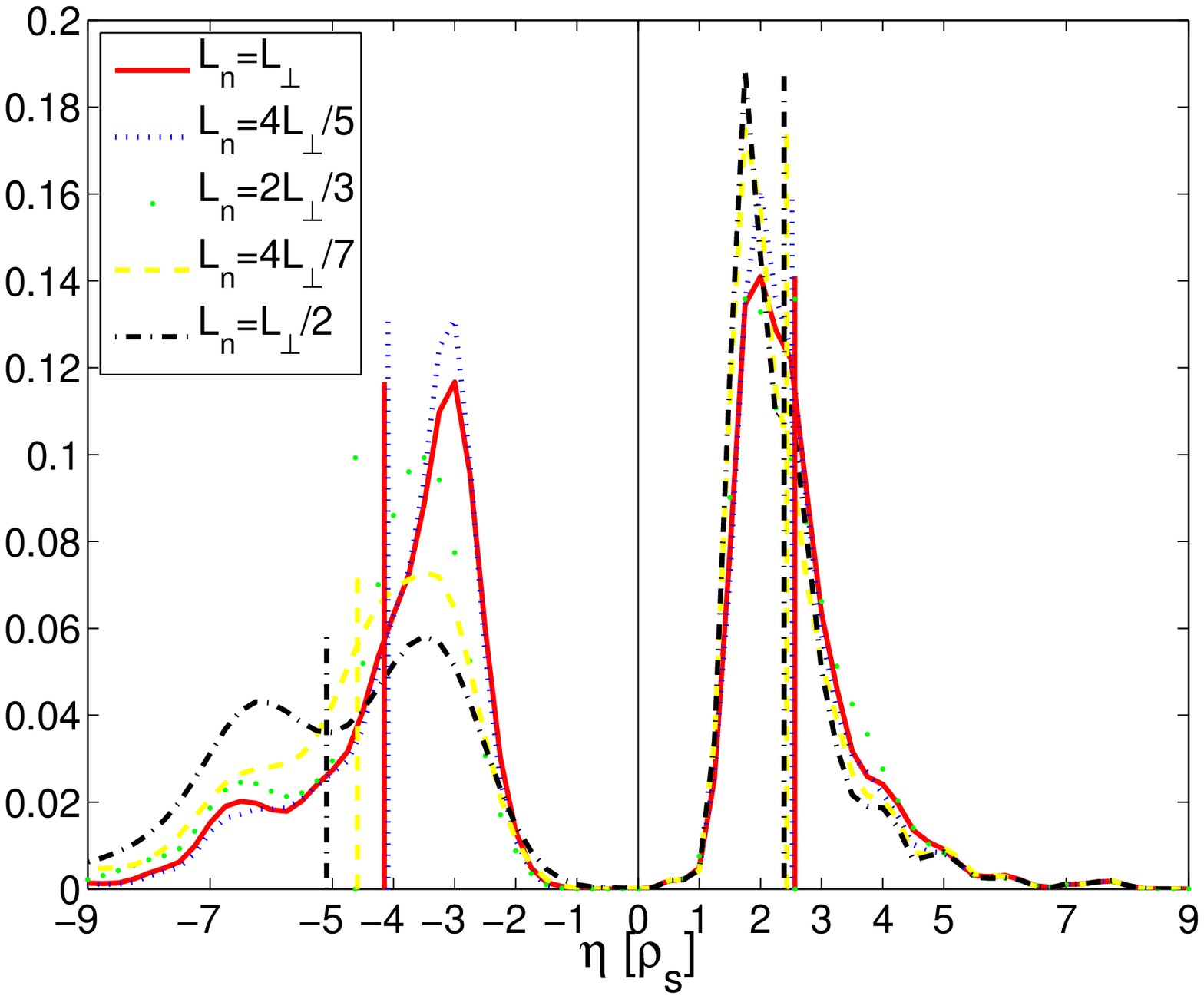} 
(a-3)~\includegraphics[width=4.2cm,height=4.3cm]{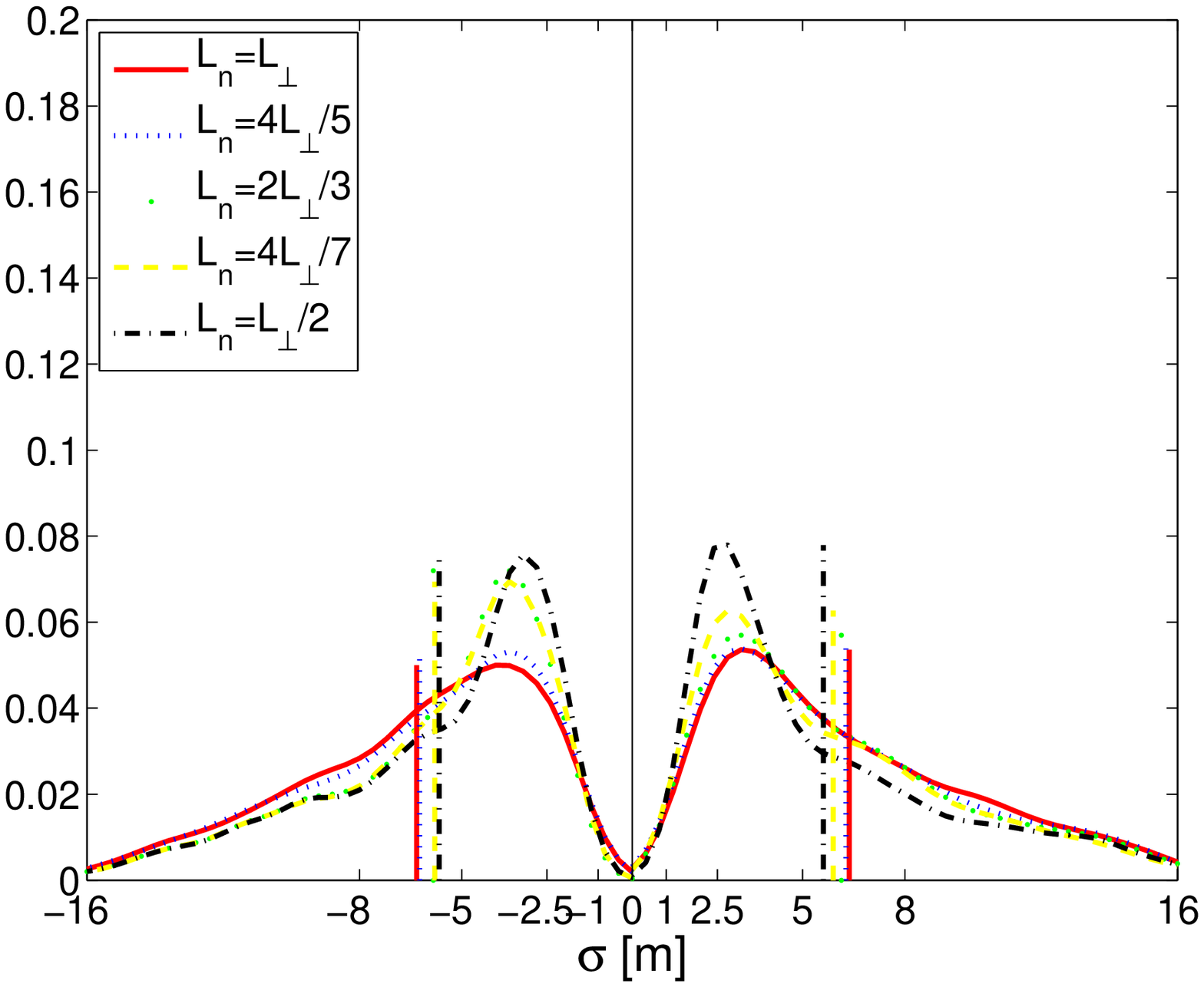}\\
(b-1)~\includegraphics[width=4.2cm,height=4.3cm]{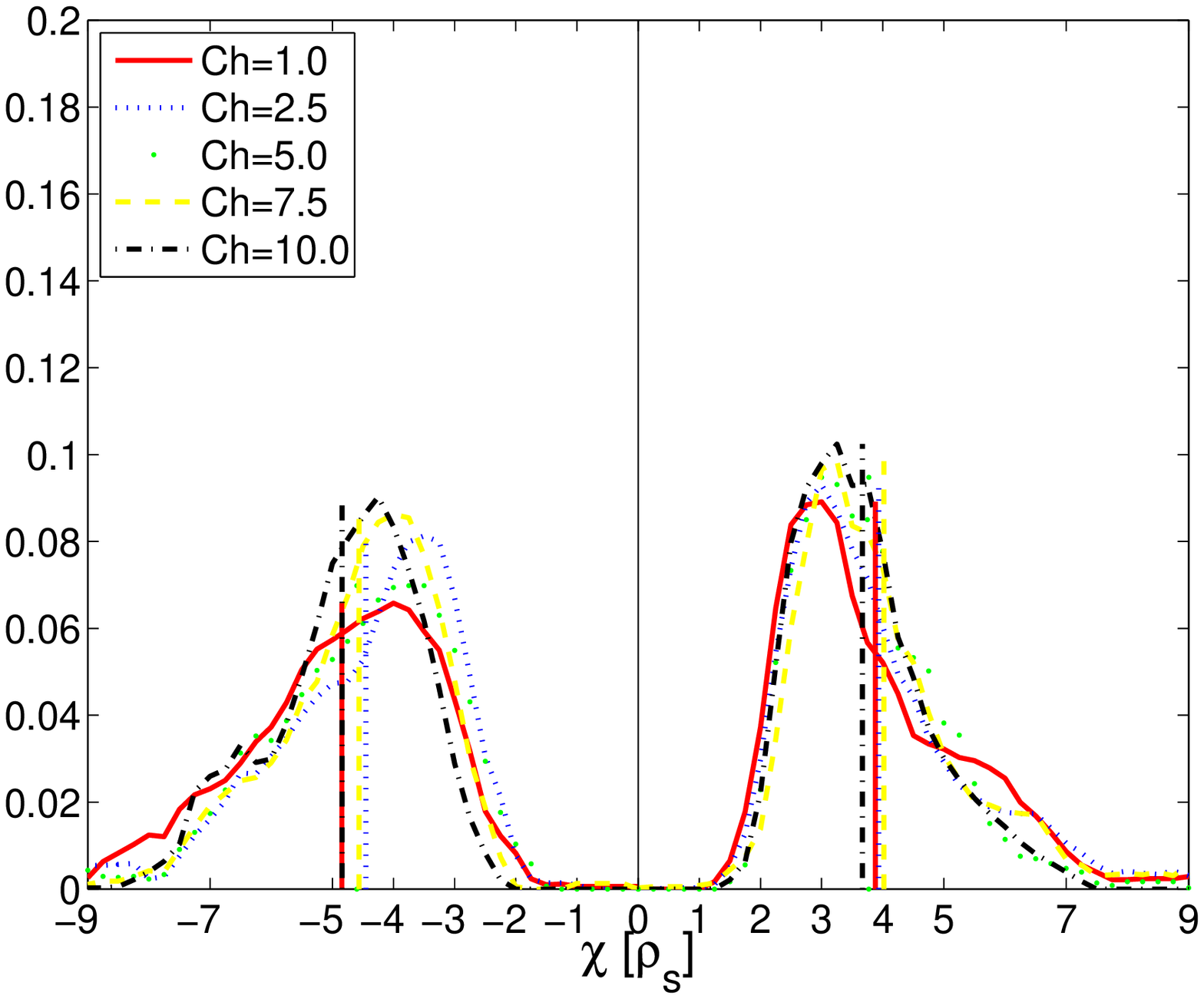}
(b-2)~\includegraphics[width=4.2cm,height=4.3cm]{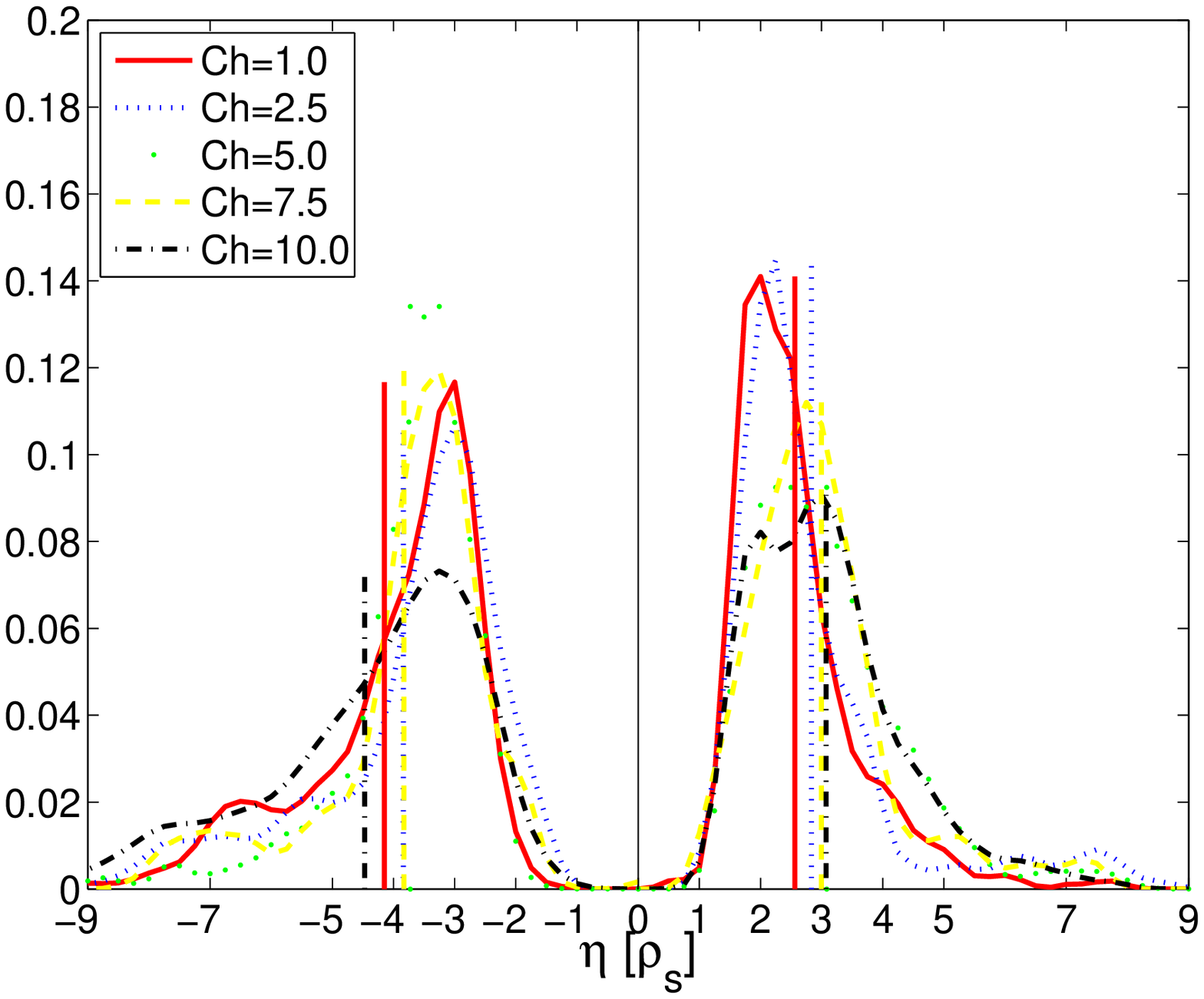}
(b-3)~\includegraphics[width=4.2cm,height=4.3cm]{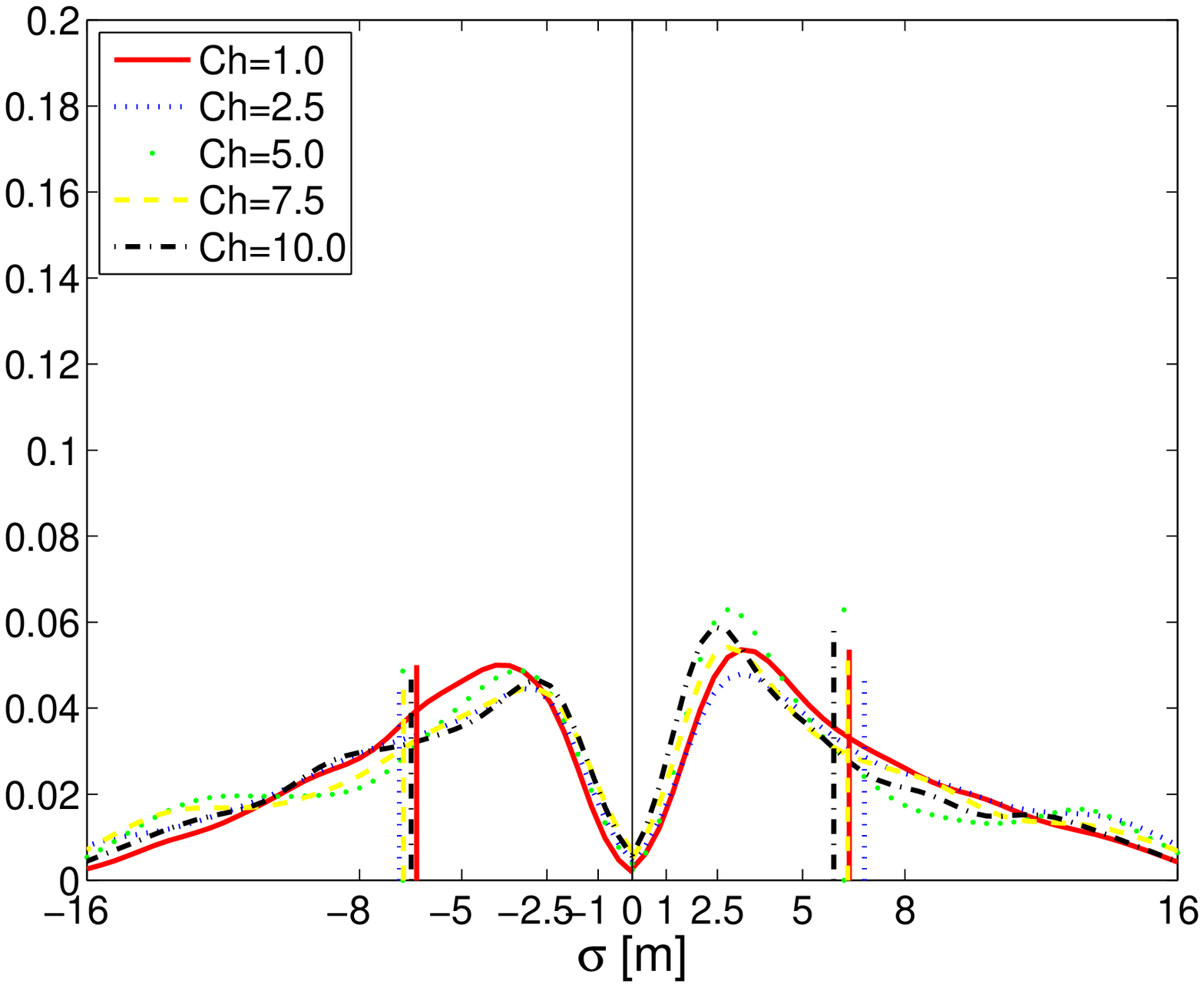}\\
(c-1)~\includegraphics[width=4.2cm,height=4.3cm]{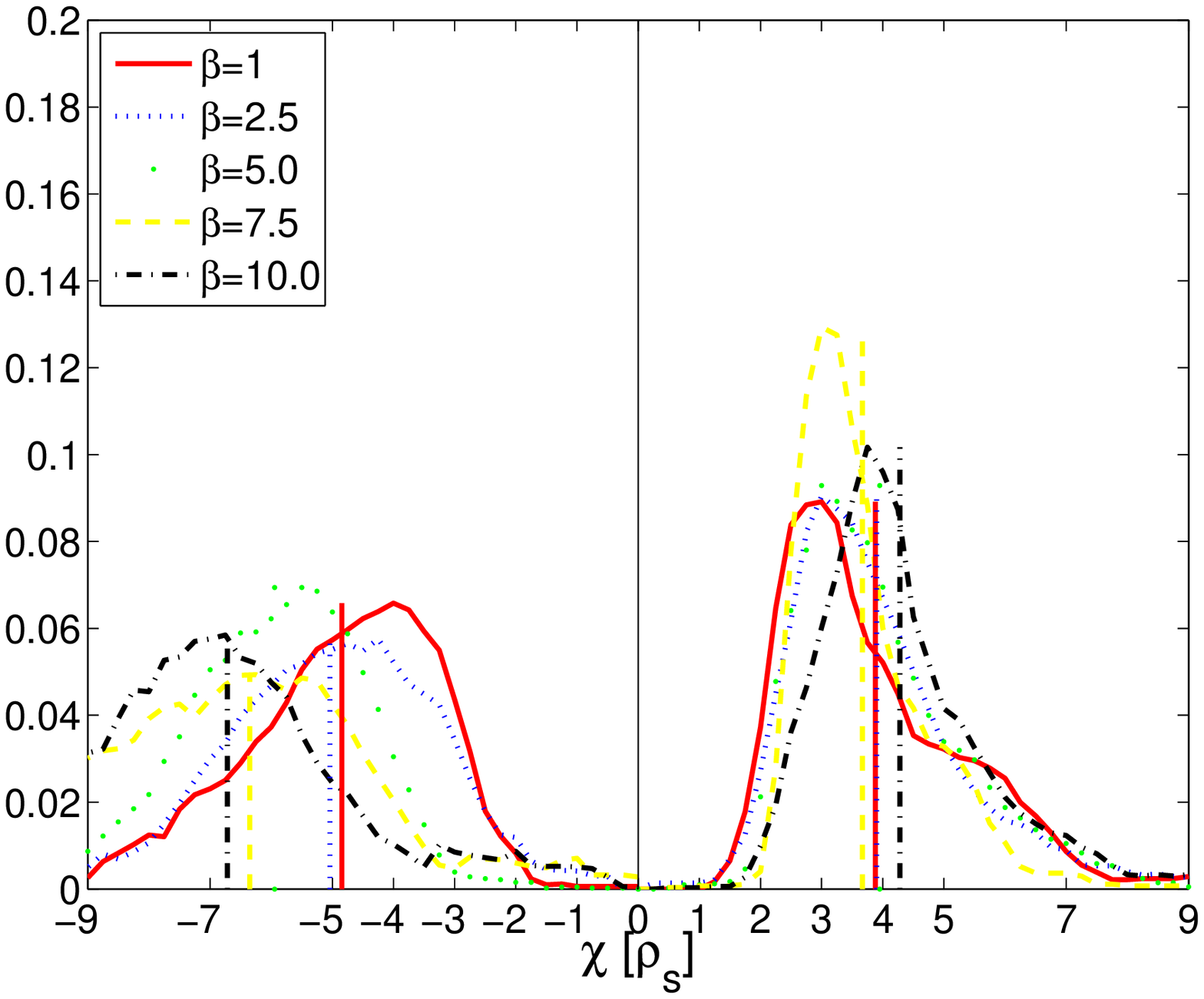}
(c-2)~\includegraphics[width=4.2cm,height=4.3cm]{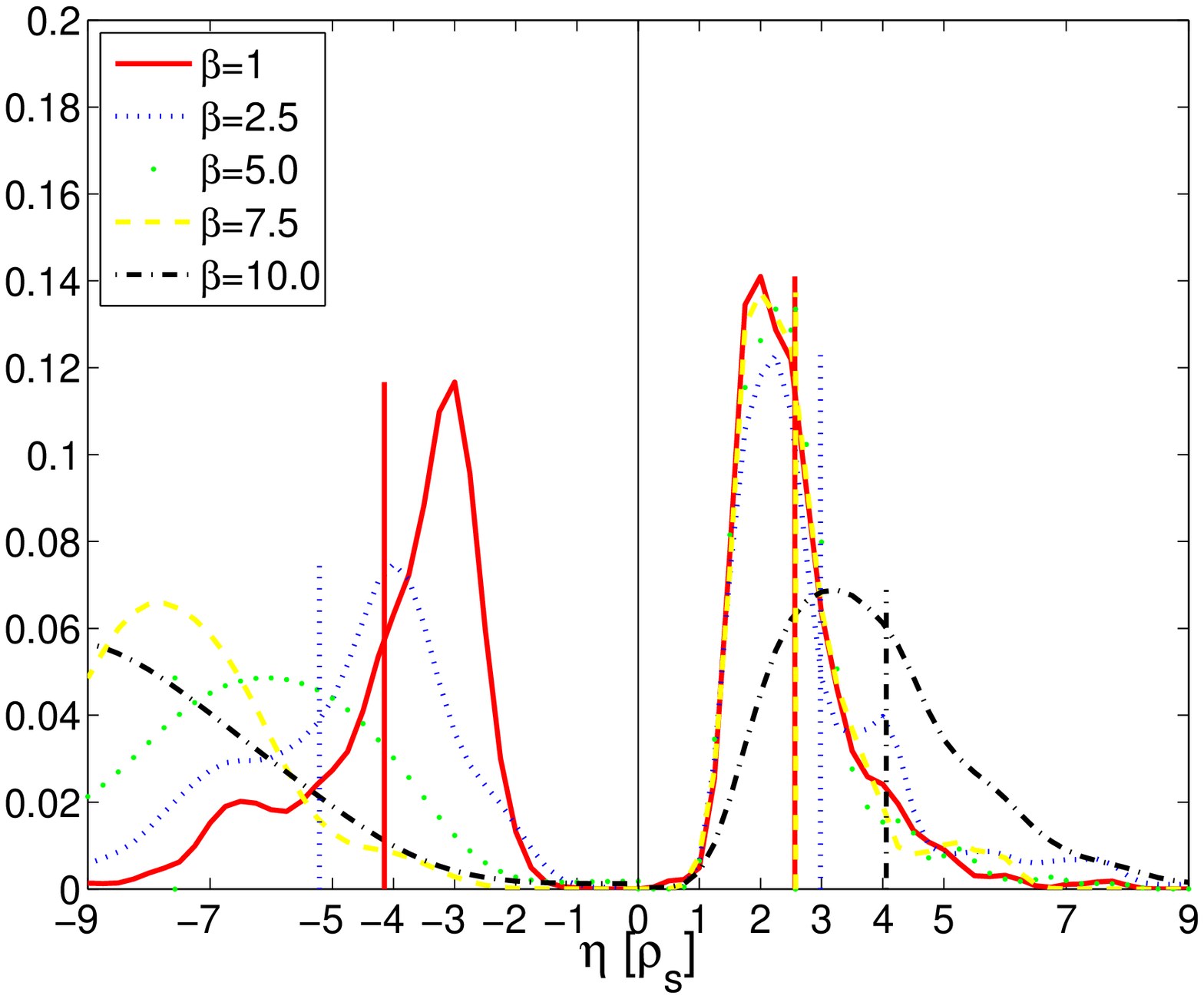}
(c-3)~\includegraphics[width=4.2cm,height=4.3cm]{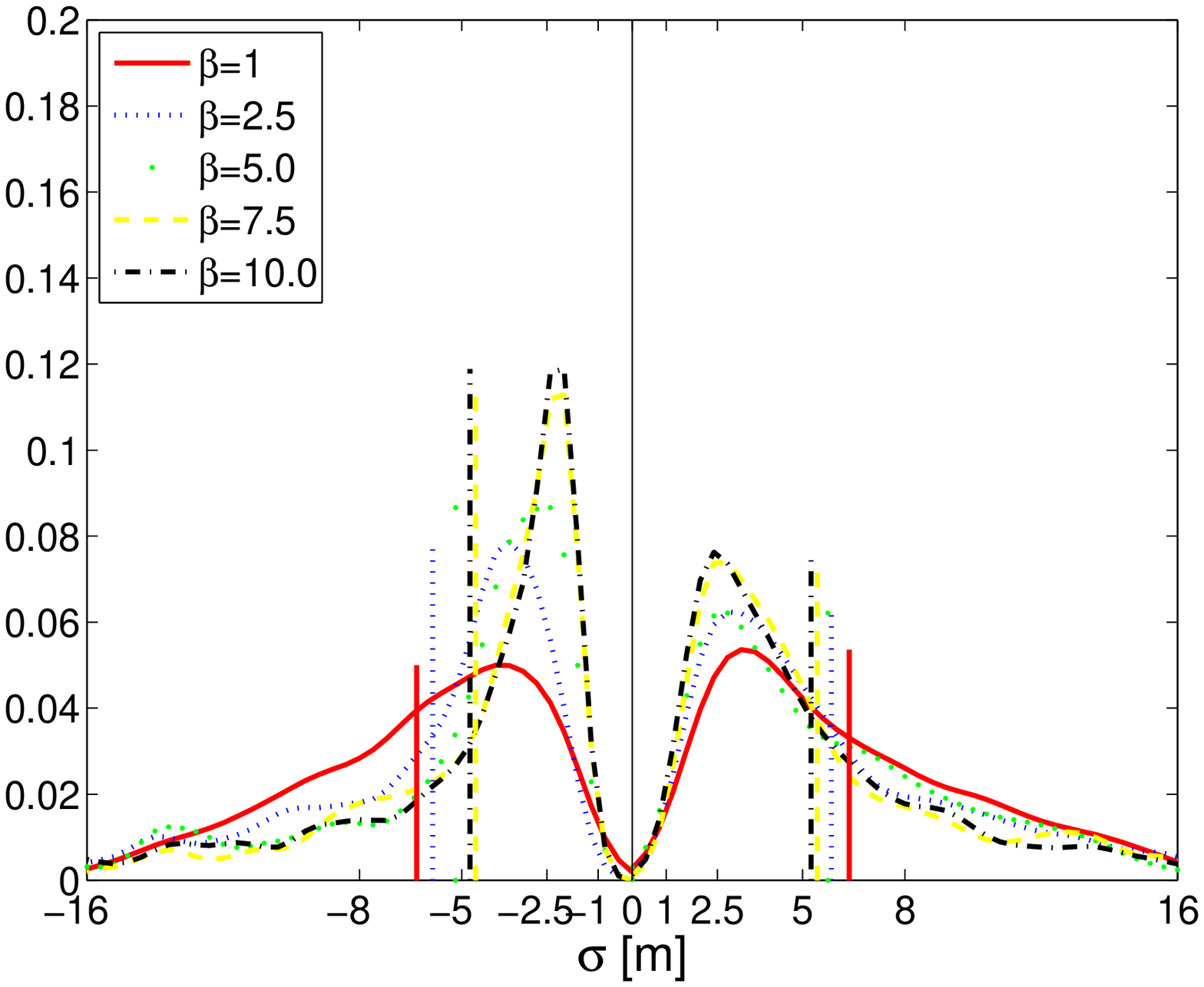}\\
(d-1)~\includegraphics[width=4.2cm,height=4.3cm]{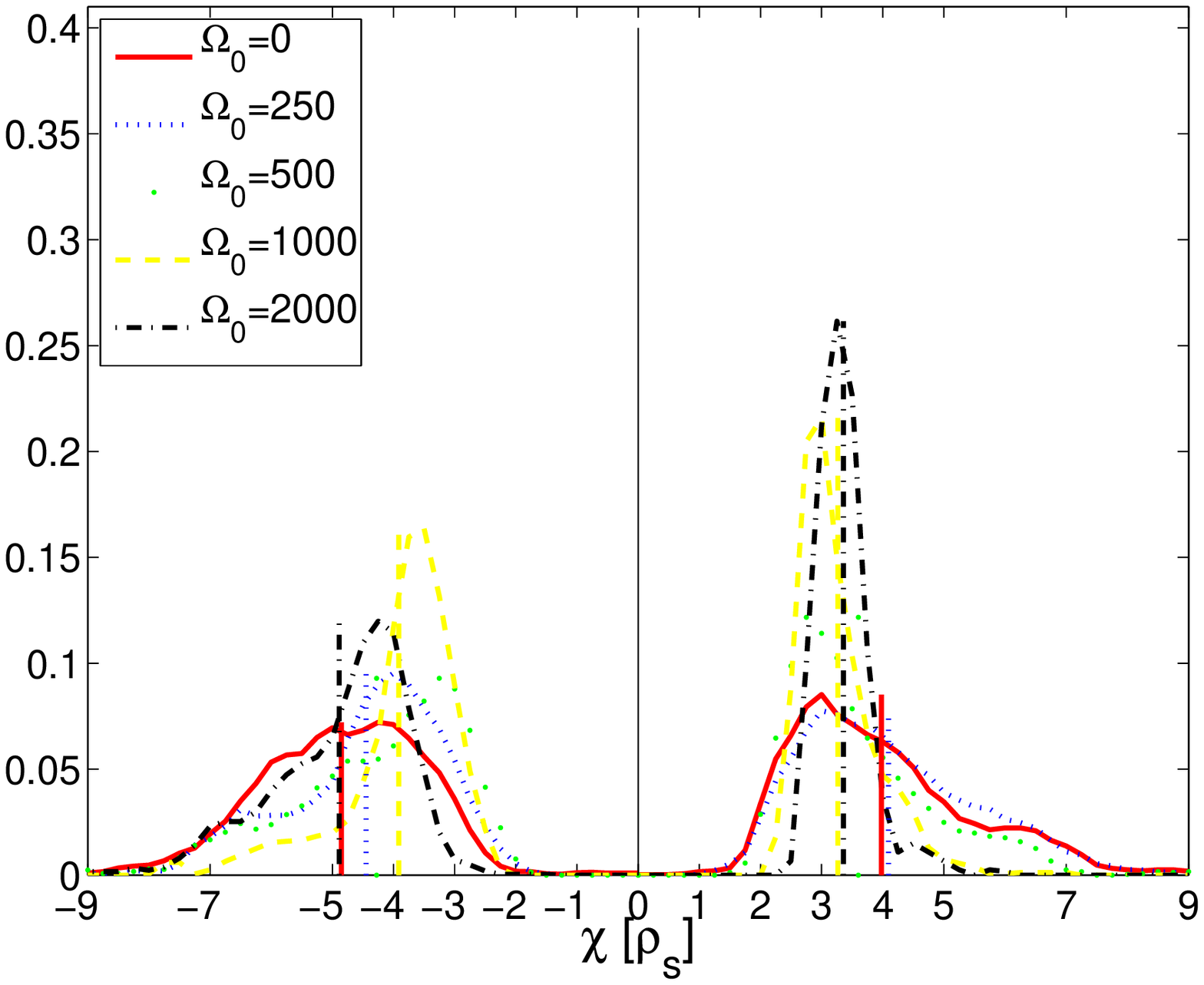}
(d-2)~\includegraphics[width=4.2cm,height=4.3cm]{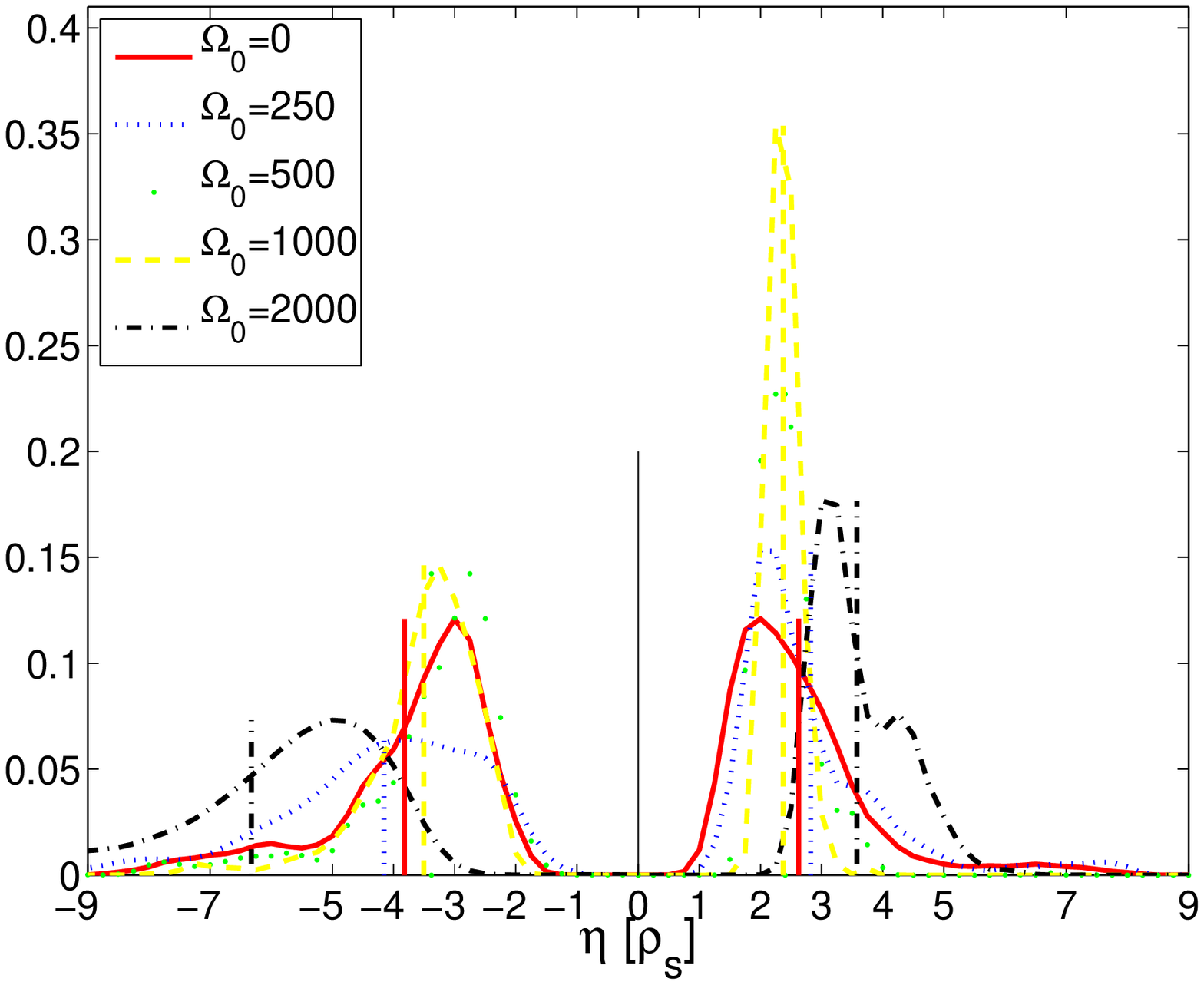}
(d-3)~\includegraphics[width=4.2cm,height=4.3cm]{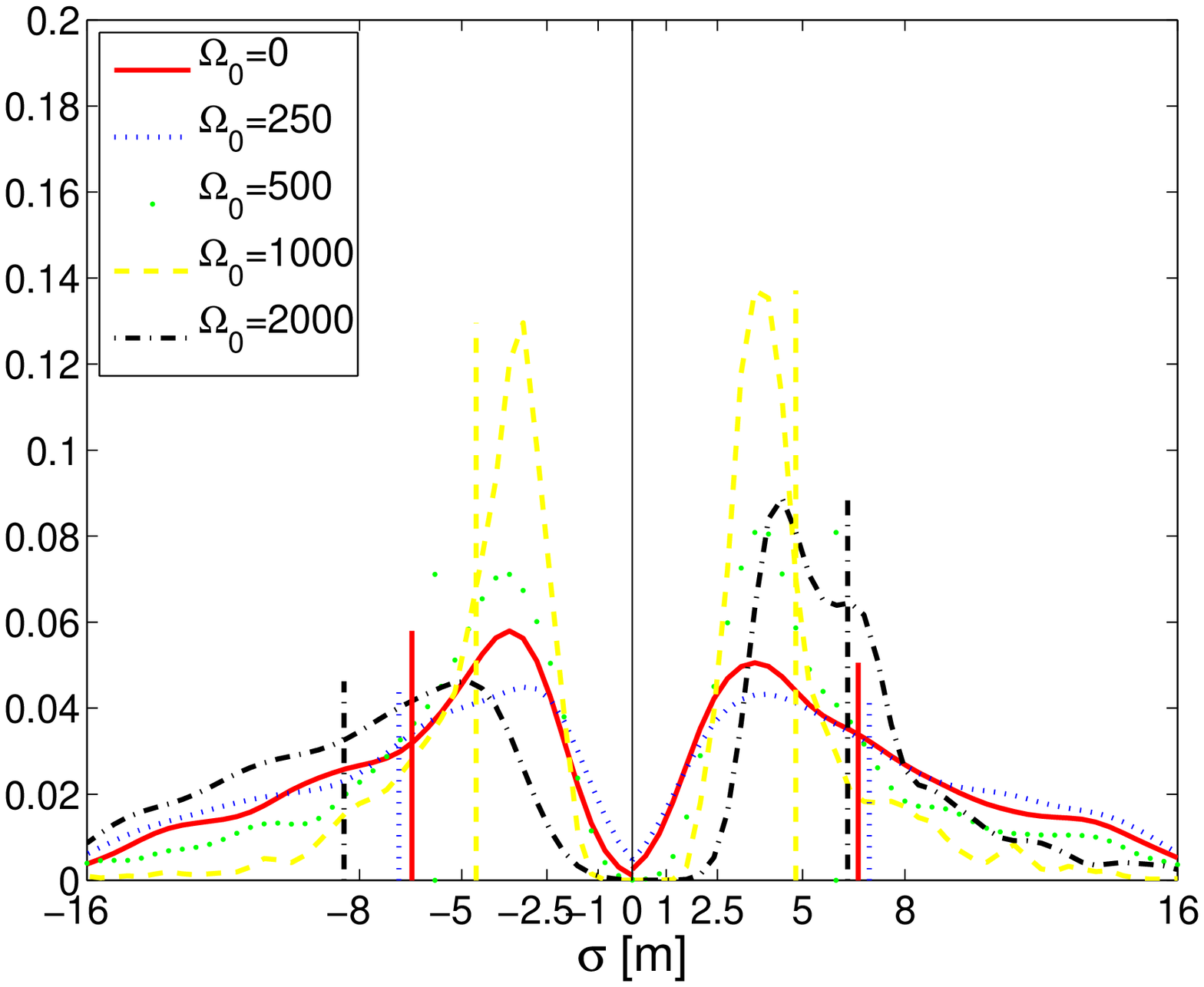}
\caption{\sl Correlation length PDFs scalings with (a) $L_{n}$, (b) $\hat C$, (c) $\hat
  \beta$, (d) $\Omega_0$: (1) radial, (2) poloidal (perpendicular), (3) parallel direction.
\vspace{2cm}}
\label{fig:spatialcorr}
\end{figure}

For the correlation length analysis fluctuations of density $n$ and
electrostatic potential $\phi$ have been recorded at three further positions, once
radially separated by $\Delta x$ , once poloidally (or rather, perpendicular
to radius and magnetic field in the drift plane) by $\Delta y$, and also along
the magnetic field line by $\Delta z$. 
Correlation length PDFs $P(\lambda_x)$, $P(\lambda_y)$, $P(\lambda_z)$,
as defined in eq.~(\ref{eq:pdfcorrlength}), are shown in
fig.~(\ref{fig:spatialcorr} a) for increasing density gradient $L_{n}$.

The potential correlation lengths, drawn as PDFs on the negative axis
of plots (\ref{fig:spatialcorr} a), show a slight rise in
correlation lengths radially and poloidally, whereas the correlation length
along the magnetic field line is slightly reduced when $L_{n}$ is increased. 
The density PDFs (on the positive axis) show an increase in events with short
spatial correlations, both radially and in parallel, whereas for larger scales
the probability is reduced. The density PDFs in poloidal direction remain
nearly unchanged.

All spatial correlation length PDF appear to be far from Gaussian, with a steep
peak at small scales and long tails for larger correlation lengths. 
The perpendicular scales are consistent with dominant turbulent vortex
structures of the order of a few $\rho_s k_{\perp}$.
The net effect from an increased density gradient in the constant in general
are slightly smaller spatial correlation lengths for density fluctuations.  

%%%%%%% FIGURE 3: spatial corr PDFs for Ln:

\subsection{Collisionality scaling}

Next, the effect of reduced collisionality on edge turbulence correlations is analyzed.
The collisionality $\hat C \propto 1/T_{e}$ scales inversely with the
electron temperature $T_{e}$. 
In table (\ref{table::parameters_Ch}) the parameters used for this simulation 
series are summarised. A collisionality parameter $\hat C = 1$ corresponds
to a roughly doubled temperature compared to $\hat C = 10$.

In fig.~(\ref{plot:energetics} b) the energetics scaling with collisionality is shown. 
Except for the zonal flow energy $\langle v_{ZF}^{2} \rangle$ all global
energy averages increase linearly with collisionality. 
For the radial density transport $\Gamma_x$ not only the mean value, but the
fluctuation width (drawn as vertical deviation bars around the mean) increases. 

In fig.~(\ref{fig:tempcorr} b) AC times of density and potential
fluctuations are shown. Mean values of AC times, drawn  as
vertical lines in fig.~(\ref{fig:tempcorr} b-1), show no clear scaling with collisionality. 
Only the $E \times B$ AC time, on the negative half in 
fig.~(\ref{fig:tempcorr} b-2), shows a clear decrease for increasing $\hat{C}$.

All time scales are as usual normalized to the drift drift time scale $L_{\perp}/c_s$.
In this series of simulations, however, $L_{\perp}/c_s$ is varied (according
to table~\ref{table::parameters_Ch}) in addition to  $\hat C$. 
The reverse scaling of $\tau_{E \times B}$ with $T_e$, which is evident in
fig.~(\ref{fig:tempcorr} b-2) in this $T_e$-dependent normalization, would
also be qualitatively preserved if $\tau_{E \times B}$ were plotted in
physical units (with the $T_e$ dependent normalization eliminated).

Spatial correlation lengths, shown in fig.~(\ref{fig:spatialcorr} b), do
not reveal any clear change in the correlation of fluctuation signals with
collisionality.  

\begin{table}
%\begin{center}
\begin{tabular}{llllll} 
Sim. & $\hat{C}$ & $T_{e}$ & $n_{0}$ &$L_{\perp}/c_{s}$ & $\rho_{s}$ \\
no. &  & $[eV]$ & $[10^{18}\; m^{-3}]$ & $[10^{-6}\; s]$ & $[10^{-3} \; m]$ \\

\hline
% \\
1 &  1.0 & 51.8 & $5.56$ & $0.80$  & 1.00\\
2 &  2.5 & 38.2 & $7.60$ & $0.93$ & 8.93 \\
3 &  5.0 & 30.3 & $9.51$ & $1.00$ & 7.95 \\
4 &  7.5 & 26.5 & $10.9$ & $1.12$ & 7.43 \\
5 &  10.0 & 24.0 & $12.0$ & $1.18$ & 7.10 \\
\hline
\end{tabular}
\caption{Parameters for simulations with increasing collisionality $\hat{C}$} 
\label{table::parameters_Ch} 
%\end{center}
\end{table}

\subsection{Plasma beta scaling}

In the H-mode the plasma pressure in the pedestal is elevated.  For constant
magnetic field strength, the plasma beta $\beta=n^{2}T^{2}/\mu_{0}^{2}/B^{2}$ rises. 
This motivates the following simulation series, where the magnetic beta is
increased, while keeping the collisionality constant. 
The variation of the electron temperature $T_{e}$, the particle density $n$,
the time and space scales for the various simulation runs are listed in table
(\ref{table::parameters_bh}). 

The global energetics are shown in fig.~(\ref{plot:energetics} c). 
A reduction of zonal flows (drawn as solid line) to about a third for the
$\hat{\beta}=10$ compared to the reference parameter case ($\hat \beta =1$), is caused
by the enhanced Maxwell stress \cite{Naulin05}.  

The zonal flow shear, the density free energy, the $E \times B$ flow energy as
well as the radial $E \times B$ density transport increase with beta. 

In fig.~(\ref{fig:tempcorr} c-1) the AC PDFs $P(\tau_{AC}(n))$, drawn
in the positive half space, show a slight decrease of $\tau_{AC}=3~L_{\perp}/c_{s}$
down to $\tau_{AC}=2~L_{\perp}/c_{s}$ for beta rising from $1$ to $10$.

Comparison with the $E \times B$ and convective timescales on the negative and
positive sides of fig.~(\ref{fig:tempcorr} c-2) 
shows that the lowered AC time for density fluctuations 
is accompanied by an an increasing turbulent $E \times B$ drift velocity. 
The convective time scale shifts to larger values for rising magnetic beta.

The perpendicular spatial correlation length PDFs for the potential,
$P(\lambda_x)$ and $P(\lambda_y)$ in fig.~(\ref{fig:spatialcorr} c-1,2), show
growing tails (negative axis).  
The perpendicular size of potential structures grows with beta. 
Together with an increasing fluctuation energy  
this results in
nearly unchanged mean decorrelation times $\tau_{AC}$, refered from the mean
of $P(\tau_{AC})$, drawn on the negative half-space of figure (\ref{fig:tempcorr} c-1). 

The size of density perturbations increasingly differs from the potential perturbations
due to the higher non-adiabaticity (via magnetic flutter) of the electrons. 
The result is an average density correlation length of 
$\lambda_x \sim \lambda_y \sim 4 \rho_s$ 
for the density, and 
$\lambda_x \sim 7 \rho_s$ and $\lambda_y \sim 6 \rho_s$ 
for the potential at $\hat{\beta}=10.0$. 

Along magnetic field lines $P(\lambda_z)$ drawn in fig.~(\ref{fig:spatialcorr} c-3)
shows a clear decrease in correlation lengths of both density and potential,
caused by the enhanced magnetic flutter.

\begin{table}
%\begin{center}
\begin{tabular}{llllll} 
Sim. & $\hat{\beta}$ & $T_{e}$ & $n_{0}$ &$L_{\perp}/c_{s}$ & $\rho_{s}$ \\
no. &  & $[eV]$ & $[10^{18} \; m^{-3}]$ & $[10^{-6} \; s]$ & $[10^{-3} \; m]$ 
\\
\hline
1 &  1.0 & 52 & 5.56 & 0.80  & 1.00\\
2 &  2.5 & 70 & 10.2 & 0.69  & 1.21 \\
3 &  5.0 & 88 & 16.2 & 0.61  & 1.36\\
4 &  7.5 & 101 & 21.3 & 0.57 &  1.45 \\
5 &  10.0 & 111 & 25.8 & 0.55 &  1.52 \\
\hline
\end{tabular}
\caption{Parameters for simulations with increasing magnetic $\hat{\beta}$} 
\label{table::parameters_bh}
%\end{center}
\end{table}

\subsection{Imposed shear flow scaling}

Further, a simulation series has been performed with the aim of analysing the
impact of an imposed sheared $E \times B$ mean flow on correlations. 

An external electrostatic zonal potential field $\phi_0(x)$ is applied through
the nonlinear advection operators as  $[\phi,f] \rightarrow [\phi+\phi_0,f]$: 
\begin{eqnarray}
\phi_{0}(x) & = & (1/2) \Omega_{0} \left(x+L_x/2 \right)^{2} \\
v_{0} (x) & = & \Omega_{0} \left(x+L_x/2 \right)  \\ 
\partial_x v_{0}(x)  & = & \Omega_{0} 
% 2 \Phi_{0} \left(1 \over L_{x}^{2}\right)  
\end{eqnarray}

The application of $\phi_{0}$ results in a radially increasing
$E \times B$ mean drift flow with velocity $v_{0}$
and constant flow shear $\Omega_0 = (0, 250, 500, 1000, 2000)$.

In fig.~(\ref{plot:energetics} d) the turbulent transport and
fluctuation amplitudes show a reduction for all levels of an imposed flow shear.
The zonal flow amplitude increases strongly for moderate $ \Omega_{0} = 250 - 500$ due
to enhanced zonal vorticity coupling $v_0^2 \sim \langle \Omega \rangle \langle
R_e \rangle$ of the Reynolds stress drive.   
For larger $ \Omega_{0}$ the zonal flows appear strongly reduced, when the
Reynolds stress $R_e = \partial_x \phi \; \partial_y \phi$ is lowered by the
quenched fluctuation amplitudes.

The fluctuation AC time PDF for the shear flow experiment in
fig.~(\ref{fig:tempcorr} d-1) 
shows for $\Omega_{0}=250-500$ a shift of the maximum to higher AC times
(around $3-4~L_{\perp}/c_{s}$) for both density and potential fluctuations. 
For higher values of $\Omega_{0}$ the maximum of the PDF shifts to smaller
$\tau_{AC}$, and peaks sharply for the density along the positive axis of plot
(\ref{fig:tempcorr} d-1).  
Vertical lines indicate mean values $\langle \tau_{AC} \rangle_{t}$, which
reflect first the trend to longer self correlation times and then a drop to
shorter living structures. 

To properly reflect the lifetime of a turbulent structure, the AC statistics
should be computed from a fluctuation time series taken in a co-moving
frame, or with the mean flow velocity subtracted from the field. 
At a fixed probe position (which has been applied here for consistency with
experimental measurements) the AC statistics not only maps the eddie turnover, but also
the background advection of the structure. The drop in the AC PDF in
fig.~(\ref{fig:tempcorr} d-2) is therefore partly debted to the faster decay
of perturbations, but also to the higher convective $E \times B$ velocity of convection.

Characteristic $E \times B$ convection times $\tau_{E \times B}$ 
drawn as PDFs in the left half space of fig.~(\ref{fig:tempcorr} d-2)
increase for growing $\Omega_{0}$. This suggests that 
convection is mainly caused by the fixed background $E \times B$ mean flow,
whereas the convection caused by $E \times B$ drift fluctuations are
suppressed by the mean flow. Along the positive axis the PDFs flatten for
increasing $\Omega_{0}$. 

Spatial correlation PDFs are shown in fig.~(\ref{fig:spatialcorr} d).
For $\Omega_{0} = 250 -500$, the perpendicular functions $P(\lambda_x)$, and
$P(\lambda_y)$ do not change significantly. 
Further increasing  $\Omega_{0}$ gives a shift of radial correlation lengths
to smaller values, for density from around $4~\rho_{s}$ to $3~\rho_{s}$. 

The poloidal correlation function $P(\lambda_y)(n)$ on the right half space of
fig.~(\ref{fig:spatialcorr} d-2) increases to around $3.5~\rho_{s}$ for
maximal imposed flow strength, and the potential correlation PDF along the
negative axis show an increase to $6~\rho_{s}$. 
PDFs in fig.~(\ref{fig:spatialcorr} d-1,2) of perpendicular correlation
lengths show that the average length for $\Omega_{0}=0$ is larger radially
than poloidally for the density. 
For the imposed shear flow of amplitude $\Omega_{0}=2000$ perturbations of
density as well as of $\phi$ are strained out poloidally and radially quenched.

Fig.~(\ref{fig:spatialcorr} d-3) shows a slight reduction of parallel correlation length
PDFs $P(\lambda_z)$ with initially rising flow shear amplitude for both potential and density . 
For the largest values of the imposed flow shear (drawn as black dash dotted
line) the parallel correlation length of $\phi$ increases from $6~m$ to about $8~m$.
The density PDFs $P(\lambda_z)(n)$ shows only a damping of events with a very low
parallel correlation length $\lambda_z<2.5~m$.

\newpage
\section{Conclusions} 

Correlation lengths of density and electrostatic potential fluctuations for
conditions relevant to an L-mode tokamak edge plasma near the L-H transition
have been analysed by numerical simulation of drift-Alfv\'en turbulence. 
Five parameter scalings have been independently performed: density gradient length $L_n$,
the collisionality $\hat C$, the plasma beta $\hat \beta$, and by imposing a
flow shear $\Omega_0$.

A reduction of $L_n$ (corresponding to a pedestal profile steepening)
results in slightly enhanced perpendicular correlations lengths for $\phi$,
and reduced correlation lengths along the magnetic field lines. 

Reducing collisionality $\hat C$ (corresponding to a rise in pedestal temperature) 
did not show any clear and significant scaling of all correlation lengths. 

Increasing the plasma beta $\hat \beta$ has different effects on density and
potential correlations. Perpendicular correlation lengths for $\phi$ increase,
whereas the density correlation length PDF is shifted to smaller spatial scales. 
Parallel correlation lengths are reduced for both $n$ and $\phi$.

Externally imposing a flow shear $\Omega_0$ was found to significantly enhance 
poloidal and parallel correlations lengths of $\phi$ only for very strong
shearing rates. Density correlation lengths are increased poloidally but are
reduced along magnetic field lines. Radially a reduction of correlation
amplitudes of $n$ and $\phi$ has been found.

It can be concluded from these parameter scaling simulations, that
experimentally observed long-range correlations near or at the transition to
H-mode states are likely caused by the straining
effect of a strongly sheared flow on the turbulence. 
The zonal flows are amplified strongly only for moderate imposed
mean flow shear, whereas long-range correlations appear only for strong
external shearing. 
All other plasma parameters scalings, which appear towards a
transition, have either a weak or reducing influence on correlation lengths.

%..................................................................
\newpage
\section*{Acknowledgements}

This work was partly supported by the Austrian Science Fund (FWF) project 
no.~Y398, by a junior research grant from University of
Innsbruck, and by the European Commission under the Contract of Association
between EURATOM and \"OAW carried out within the framework of the European
Fusion Development Agreement (EFDA). The views and opinions expressed herein
do not necessarily reflect those of the European Commission.

%..................................................................

\end{document}